%% file: ms.tex
\documentclass[apj,tighten,iop]{emulateapj2}

\input{defs}

\begin{document}
\title{PRIMUS: One- and Two-Halo Galactic Conformity at $0.2<\,$\MakeLowercase{\emph{z}}$\,<1$}
\shorttitle{PRIMUS: Galactic Conformity}
\shortauthors{Berti et al.}

\author{Angela M. Berti\altaffilmark{1},
	Alison L. Coil\altaffilmark{1},
	Peter S. Behroozi\altaffilmark{2,3},
	Daniel J. Eisenstein\altaffilmark{4},
	Aaron D. Bray\altaffilmark{4},
	Richard J. Cool\altaffilmark{5}, and
	John Moustakas\altaffilmark{6}
}

\altaffiltext{1}{Center for Astrophysics and Space Sciences, Department of Physics, University of California, 9500 Gilman Dr., La Jolla, San Diego, CA 92093, USA}
\altaffiltext{2}{Department of Astronomy, University of California at Berkeley, Berkeley, CA 94720, USA} 
\altaffiltext{3}{Hubble Fellow} 
\altaffiltext{4}{Harvard-Smithsonian Center for Astrophysics, 60 Garden Street, Cambridge, MA 02138, USA} 
\altaffiltext{5}{MMT Observatory, Tucson, AZ 85721, USA} 
\altaffiltext{6}{Department of Physics and Astronomy, Siena College, 515 Loudon Road, Loudonville, NY 12211, USA} 

\begin{abstract}

We test for galactic conformity at $0.2<z<1.0$ to a projected distance of 
5~Mpc using spectroscopic redshifts from the PRism MUlti-object Survey (PRIMUS).
Our sample consists of $\sim60,000$ galaxies in five separate fields covering 
a total of $\sim5.5$ square degrees, which allows us to account for cosmic 
variance.
We identify star-forming and quiescent ``isolated primary" (i.e., central) 
galaxies using isolation criteria and cuts in specific star formation rate.
We match the redshift and stellar mass distributions of these samples, to 
control for correlations between quiescent fraction and redshift and stellar 
mass.
We detect a significant $(>3\sigma)$ one-halo conformity signal, or an 
excess of star-forming neighbors around star-forming central galaxies, 
of $\sim5$\% 
on scales of 0--1~Mpc and a $2.5\sigma$ two-halo signal of $\sim1$\% on scales 
of 1--3~Mpc.
These signals are weaker than those detected in SDSS and are 
consistent with galactic conformity being the result of large-scale tidal fields 
and reflecting assembly bias.
We also measure the star-forming fraction of central galaxies at fixed stellar 
mass as a function of large-scale environment and find that central galaxies 
are more likely to be quenched in overdense environments, independent of 
stellar mass.
However, we find that environment does \emph{not} affect the star formation 
efficiency of central galaxies, as long as they are forming stars.
We test for redshift and stellar mass dependence of the conformity signal 
within our sample and show that large volumes and multiple fields are 
required at intermediate redshift to adequately account for cosmic variance.

\end{abstract}

\input{intro}

\input{data}

\input{results}

\input{discussion}

\input{conclusion}

\acknowledgements
We gratefully acknowledge Andrew Hearin for thoughtful feedback on a draft of this paper.
We also thank the CFHTLS, COSMOS, DLS, and SWIRE teams for their public data releases and/or access to early releases.
This paper includes data gathered with the 6.5 m Magellan Telescopes located at Las Campanas Observatory, Chile.
We thank the support staff at LCO for their help during our observations, and we acknowledge the use of community access through NOAO observing time.
Some of the data used for this project are from the CFHTLS public data release, which includes observations obtained with MegaPrime/MegaCam, a joint project of CFHT and CEA/DAPNIA, at the Canada-France-Hawaii Telescope (CFHT) which is operated by the National Research Council (NRC) of Canada, the Institut National des Science de l'Univers of the Centre National de la Recherche Scientifique (CNRS) of France, and the University of Hawaii.
This work is based in part on data products produced at TERAPIX and the Canadian Astronomy Data Centre as part of the Canada-France-Hawaii Telescope Legacy Survey, a collaborative project of NRC and CNRS.

Funding for PRIMUS has been provided by NSF grants AST-0607701, 0908246, 0908442, 0908354, and NASA grant 08-ADP08-0019.
A.L.C. acknowledges support from the NSF CAREER award AST-1055081.

\bibliography{refs}

\end{document}

%% file: defs.tex
\usepackage{epstopdf}

\usepackage{natbib}
\usepackage[usenames]{color}
\definecolor{red}{rgb}{0.75,0,0}
\definecolor{green}{rgb}{0,0.5,0}
\definecolor{blue}{rgb}{0,0,0.75}
\usepackage[ colorlinks=true, 
             urlcolor=blue,
             filecolor=blue,
             linkcolor=red,
             citecolor=blue,
             pdftitle={pdf title},
             pdfauthor={pdf author},
             pagebackref, 
             bookmarks,
             bookmarksopen=true]{hyperref}

\usepackage{lineno}	
\usepackage{xspace}    
\usepackage{microtype} 
\usepackage{amsmath}  
\usepackage{amssymb}
\usepackage{float}
\usepackage{booktabs}

\usepackage[normalem]{ulem}
\newcommand\redline{\bgroup\markoverwith{\textcolor{red}{\rule[0.5ex]{2pt}{0.4pt}}}\ULon}

\usepackage{makecell} 
\usepackage{rotating} 

\newcommand{\Rproj}{\ensuremath{\mathcal{R}_{\text{proj}}}}

\newcommand{\degsq}{\ensuremath{\mathrm{deg}^2}\xspace}

\newcommand{\msun}{\mathcal{M}_{\sun}\xspace}
\newcommand{\signorm}{\ensuremath{\xi_{\textrm{norm}}}\xspace}
\newcommand{\flate}{\ensuremath{f_{\textrm{SF}}}\xspace}

\newcommand{\mstar}{\ensuremath{\mathcal{M}_*}\xspace}

\newcommand{\mlim}{\ensuremath{\mathcal{M_{\text{lim}}}}\xspace}
\newcommand{\mmax}{\ensuremath{\mathcal{M_{\text{max}}}}\xspace}

\newcommand{\mIP}{\ensuremath{\mathcal{M_{\text{IP}}}}\xspace}
\newcommand{\mneigh}{\ensuremath{\mathcal{M_{\text{neigh}}}}\xspace}
\newcommand{\Nneigh}{\ensuremath{N_{\text{neigh}}}\xspace}
\newcommand{\mcentral}{\ensuremath{\mathcal{M_{\text{central}}}}\xspace}

\newcommand{\kms}{km~s$^{-1}$\xspace}
\newcommand{\logM}{\ensuremath{\log\,(\mstar/\msun)}\xspace}

\newcommand{\iSEDfit}{\texttt{iSEDfit}\xspace}
\newcommand{\IDL}{\texttt{IDL}\xspace}

\newcommand{\maketmp}[1]{\expandafter\MakeUppercase\expandafter{#1}}

\newcommand{\mass}{\ensuremath{\mathcal{M}}\xspace}

\newcommand{\mhalo}{\ensuremath{\mathcal{M}_\textrm{halo}}\xspace}

\newcommand{\sfrunit}{\ensuremath{\msun\,\textrm{yr}^{-1}}}

\newcommand{\zmax}{\ensuremath{z_{\text{max}}}}

\newcommand{\sigmaBS}{\ensuremath{\sigma_{\textrm{BS}}}\xspace}
\newcommand{\sigmaJK}{\ensuremath{\sigma_{\textrm{JK}}}\xspace}
\newcommand{\sigmaz}{\ensuremath{\sigma_z}\xspace}

\newcommand{\citePB}{Be16\xspace}
\newcommand{\hubble}{\ensuremath{70~\textrm{km~s}^{-1}~\textrm{Mpc}^{-1}}}

%% file: intro.tex
\section{Introduction}\label{sec:intro}

The distributions of many related galaxy properties are bimodal, including color, star formation rate (SFR), gas fraction, and morphology
\citep[e.g.,][]{Strateva01, Kauffmann03, Baldry04, Balogh04b}.
Galaxies with lower star formation rates are usually redder and exhibit ``early-type'' morphologies, while those with higher star formation rates tend to be bluer and have ``late-type'' morphologies.
The existence of this bimodality is consistent with star formation in galaxies turning off, or quenching, rapidly, as quenching over longer timescales would result in flatter distributions of color and SFR \citep[e.g.,][]{TinkerWetzel10, Wetzel13}.

Numerous mechanisms for quenching have been proposed, including but not limited to the shock heating of infalling gas \citep*[e.g.,][]{WhiteRees78, DekelBirnboim06},
stellar and AGN feedback \citep[e.g.,][]{Croton06, Hopkins06}, and
gas heating and/or removal caused by galaxy mergers or harassment \citep[e.g.,][]{Moore96}.
Agreement on which of these mechanisms plays the largest role remains elusive in the absence of definitive evidence, although it is likely that the relative importance of these various mechanisms depends on stellar or halo mass, as well as on large-scale environment.

The recently-discovered phenomenon of galactic conformity may provide additional insights into and constraints on the quenching mechanism, and more broadly, the dependence of galaxy evolution on large-scale structure.
Galactic conformity refers to correlations between the colors and SFRs of massive central galaxies and their nearby neighboring galaxies.
It was first identified by \citet{Weinmann06} in the Sloan Digital Sky Survey \citep[SDSS;][]{York00} at $z<0.03$.
\citet{Weinmann06} identified galaxy groups in SDSS---defined as the ensemble of galaxies residing in the same dark matter halo---using a group-finding algorithm \citep{Yang05a}.
They define central galaxies as the brightest galaxy in each group, and dub all other group members satellite galaxies.
Group halo masses are estimated by assuming a correlation between group luminosity and halo mass.
\citet{Weinmann06} found that the quiescent fraction of satellite galaxies is higher for quiescent central galaxies than for star-forming central galaxies residing in halos of the same mass, and that this correlation exists for halo masses spanning three orders of magnitude, from $10^{12}$ to $10^{15}~\msun$.

\citet[][hereafter K13]{Kauffmann13} compared the specific star formation rates of central SDSS galaxies and their neighbor galaxies at fixed \emph{stellar} mass from $5\times10^9$ to $3\times10^{11}~\msun$ and found evidence of conformity at projected distances up to $\sim4$~Mpc from the central galaxy, well beyond the virial radius of a single halo.
The K13 result motivated the distinction of ``one-halo'' and ``two-halo'' conformity \citep*{Hearin15}, referring to correlations between central galaxies and their satellite galaxies within a single halo and between central galaxies and neighboring galaxies in adjacent halos, respectively.
K13 also concluded that the scale dependence of conformity is correlated with the stellar mass of the central galaxy.
Specifically, they found that two-halo conformity exists for low-mass central galaxies
(${9.7<\logM<10.3}$)
and is greatest at large separations ($>1$~Mpc).
For high-mass central galaxies (${10.7<\logM<11.3}$) the signal is confined to one-halo scales.

Galactic conformity (especially two-halo conformity) is further evidence that standard halo occupation model \citep{BerlindWeinberg02},
which presumes that the properties of a halo's galaxy population are determined solely by present-day halo mass,
does not represent the full picture of galaxy and halo clustering \citep[e.g.,][]{Kravtsov04}.
Correlations between the colors and SFRs of central galaxies and their satellites \emph{at fixed halo mass} is a clear contradiction of the assumptions of mass-only halo occupation models.

The additional dependence of halo clustering on properties beyond halo mass, such as formation epoch and large-scale environment, is referred to as \emph{assembly bias} \citep[e.g.,][]{GSW05, Wechsler06, Croton07, GaoWhite07, Zentner07, Dalal08, Tinker08, Sunayama16}.
Theoretical models provide evidence that galactic conformity may be a natural result of \emph{galaxy assembly bias}.

\citet{Hearin15} tests for two-halo conformity with three different (sub)halo abundance matching (SHAM) models of halo occupation statistics by assigning galaxies to halos in the N-body \emph{Bolshoi} simulation \citep{Klypin11}, which follows the evolution of $2048^3$ particles in a $250/h$~Mpc periodic box.
Both the standard halo occupation model, in which the quenching of central and satellite galaxies depends only on halo mass, and the delayed-then-rapid model \citep{Wetzel13}, in which satellite galaxy quenching depends on both time since accretion and halo mass at accretion time, exhibit zero two-halo conformity.
The age matching SHAM model, in which central and satellite galaxy quenching depends on halo mass and (sub)halo formation time, and the lowest SFR galaxies are assigned to the oldest halos, \emph{does} exhibit two-halo conformity comparable to that seen by K13 in SDSS.

\citet{Hearin15} also shuffles the SFRs of only satellite and only central galaxies in the age matching model, and finds that shuffling satellite galaxy SFRs has little effect on the conformity signal, while shuffling the SFRs of central galaxies erases it entirely.
This result focuses the likely connection to one between two-halo conformity and \emph{central} galaxy assembly bias.

In a follow-up paper \citet*[][hereafter He16]{Hearin16} conclude that conformity and assembly bias are alternative descriptions of the same underlying phenomenon.
Because halos that assembled earlier are more strongly clustered than more recently-assembled halos of the same mass \citep[e.g.,][]{Hahn09}, older (younger) halos inhabit more (less) dense environments and are therefore subjected to stronger (weaker) large-scale tidal fields.
Strong tidal effects inhibit the rate at which dark matter is accreted into halos, giving rise to what He16 dubs \emph{halo accretion conformity}:~the clustering of halos at fixed mass with high (lower) dark matter accretion rates.

He16 finds evidence of halo accretion conformity in the \emph{Bolshoi} simulation, and proposes that two-halo galactic conformity follows from halo accretion conformity if gas and dark matter accretion rates are sufficiently coupled \citep[e.g.,][]{WetzelNagai15}.
The same work also proposes that present-day one-halo conformity may be a direct result of two-halo conformity at higher redshift, since many satellite galaxies were their own centrals at an earlier epoch.

Additionally, He16 clearly predicts halo accretion conformity strength should diminish both with increasing redshift and with increasing halo mass, as more massive halos are less sensitive to tidal effects.
For example, for $10^{11}~\msun$ secondary halos surrounding a $10^{12}~\msun$ primary halo He16 predicts that a normalized halo accretion conformity signal at 3~Mpc of $\sim20\%$ at $z=0$ would equate to a signal of $\sim4\%$ at $z\sim1$, and to just $\sim0.5\%$ by $z\sim2$.

Both strong one- and weaker two-halo conformity have been found by \citet{Bray16a} in the hydrodynamical \emph{Illustris} simulation \citep{Vogelsberger14}.
\citet{Bray16a} also detect ``halo age conformity'' to $R\sim10$~Mpc, in which less massive old (young), ``secondary'' halos are preferentially found in the vicinity of more massive old (young), ``primary'' halos.

However, \citet{Paranjape15} argue that the two-halo conformity signal in SDSS found at fixed \emph{stellar} mass in K13 is not conclusive evidence that two-halo galactic conformity is the result of halo assembly bias.
Such a signal could also be due to one-halo conformity ``leaking'' to large scales when averaging over a range of halo masses, as scatter in the stellar mass-halo mass relation means that some galaxies with the same stellar mass inevitably reside in halos of different masses.

\citet{Kauffmann15} proposes ``pre-heating'' as an alternative explanation for galactic conformity.
In the pre-heating scenario, feedback from an early generation of accreting black holes heats gas over large scales at an early epoch, causing coherent modulation of cooling and star formation among galaxies on the same large scales.
As evidence, \citet{Kauffmann15} cites an excess number of very massive galaxies out to 2.5~Mpc around quiescent central galaxies in the same SDSS sample used in K13.
\citet{Kauffmann15} also finds that massive galaxies in the vicinity of low sSFR central galaxies at $z=0$ are 3--4 times more likely to host radio-loud active galactic nuclei (AGN) than those around a control sample of higher sSFR central galaxies.
While not explicitly stated in \citet{Kauffmann15}, if pre-heating by an early generation of AGN is responsible for two-halo conformity, the signal strength will most likely emph{increase} with redshift, which is the opposite of the He16 prediction.

Measuring a statistical effect like galactic conformity at $z>0.2$ requires very deep, relatively large-volume surveys with precise redshifts.
Not surprisingly, observational studies of conformity have until recently been limited to the redshift range of SDSS.
Searching for evidence of conformity over a much larger range of cosmic time is a valuable test of assembly bias, and may play an important role in constraining the quenching mechanism(s) at work at certain stellar masses and in certain environments.

As of this writing only a few studies have tested for conformity or a related effect at $z>0.2$.
Using photometric redshifts from three fields totaling $2.37~\degsq$
\citet{Kawinwanichakij16} test for one-halo conformity in four redshift bins over the range ${0.3 < z < 2.5}$ for central galaxies with ${\mstar > 10^{10.5}~\msun}$.
\citet{Kawinwanichakij16} estimate the average quiescent fraction of satellite galaxies in fixed apertures for stellar mass-matched samples of quiescent and star-forming central galaxies.
If we define the magnitude of a conformity signal to be the percent difference between the fraction of star-forming satellites surrounding star-forming and quiescent central galaxies, \citet{Kawinwanichakij16}
find a conformity signal of $\sim10$--30\% at ${0.6 < z < 1.6}$ on scales of $\lesssim300$ projected comoving kpc, and
a $\sim10\%$ signal at ${0.3 < z < 0.6}$.

\citet[][hereafter H15]{Hartley15} also use photometric redshifts to look for one-halo conformity in a sample of ${10^{10.5}<\mstar<10^{11.0}~\msun}$ central galaxies in the 0.77~\degsq UKIDSS UDS field at ${0.4 < z < 1.9}$.
They measure the radial density profiles of quiescent satellite galaxies
for mass-matched samples of quiescent and star-forming central galaxies.
H15 find a conformity signal of $\sim50$\% on scales of $\sim10$--350~projected kpc at $0.4<z<1.9$.

Both \citet{Campbell15} and \citet{Paranjape15} have shown that systematic error can create an artificial conformity signal.
Contamination from interlopers (galaxies not physically associated with a central galaxy that are falsely classified as satellites, or satellite galaxies falsely classified as centrals) can also bias measurements of conformity \citep[for a detailed explanation see \S2.2 of][]{Hearin15}.
Spectroscopic redshifts are therefore crucial for measuring conformity robustly.
Additionally, cosmic variance may impact a conformity signal, but the effect can be mitigated by using a large survey volume and multiple fields.
We achieve this using data from the PRIsm MUlti-object Survey \citep[PRIMUS;][]{Coil11, Cool13}.

With a survey area of $\sim9$~\degsq, a redshift precision of $\sigmaz=0.005\,(1+z)$, and four spatially-distinct fields, PRIMUS is uniquely suited for investigating one- and two-halo conformity at $0.2<z<1$.
While previous studies of conformity at $z>0.2$ have necessarily used photometric redshifts, spectroscopic redshifts allow us to much more cleanly identify isolated central-like galaxies, which is critical for a robust measurement of conformity.
PRIMUS also allows us to test the effects of cosmic variance and the need for large areas in multiple fields at intermediate redshift,
and to investigate the redshift and mass dependence of one- and two-halo conformity.

In a related forthcoming paper, Bray et al.~(2016b, in prep., hereafter Br16b) perform a complimentary analysis with cross-correlations between the PRIMUS spectroscopic and photometric galaxy samples.
Br16b measure the fraction of quiescent galaxies around all PRIMUS spectroscopic galaxies (not just central galaxies) to $\sim1/h$~Mpc.

The structure of this paper is as follows.
In \S\ref{sec:data} we describe the survey used for this study and the details of sample selection.
Our results are presented in \S\ref{sec:results}.
In \S\ref{sec:discussion} we discuss the implications of our results in the context of other conformity studies and the predictions from simulations and theory.
We summarize our findings and conclusions in \S\ref{sec:conclusion}.
Throughout this paper we assume $H_{0}=\hubble$, $\Omega_{\textrm{m}}=0.3$, and $\Omega_{\Lambda}=0.7$.

%% file: data.tex
\section{Data}\label{sec:data}

In this section we describe the PRIMUS redshift survey data, how we identify star-forming
and quiescent galaxies, and how we define the isolated primary samples used to measure 
the conformity signal within PRIMUS.

\subsection{PRIMUS}\label{sec:PRIMUS}
 
The PRIsm MUlti-Object Survey (PRIMUS) is the largest spectroscopic faint galaxy redshift survey completed to date.
The survey was conducted with the IMACS spectrograph \citep{Bigelow03} on the Magellan I Baade 6.5-meter telescope at 
Las Campanas 
Observatory, using slitmasks and a low-dispersion prism.
The design allowed for $\sim2,000$ objects per slitmask to be observed simultaneously with a spectral resolution of ${\lambda/\Delta
\lambda \sim 40}$ in a $\sim0.2~\degsq$ field of view.
Objects were targeted to a maximum depth of ${i \ge 23}$, and typically two slitmasks were observed per pointing on the sky.  
PRIMUS obtained robust redshifts \citep[${Q \ge 3}$; see][]{Cool13} for $\sim120,000$ objects at ${0<z<1.2}$ with a 
redshift precision of ${\sigmaz/(1 + z) \sim 0.005}$.

The total survey area of PRIMUS is $9.1~\degsq$ and encompasses seven distinct science fields:
the Chandra Deep Field South-SWIRE field \citep[CDFS;][]{Lonsdale03},
the 02hr and 23hr DEEP2 fields \citep{Newman13},
the COSMOS field \citep{Scoville07},
the European Large Area ISO Survey-South 1 field \citep[ES1;][]{Oliver00},
the Deep Lens Survey \citep[DLS;][]{Wittman02} F5 field,
and two spatially-adjacent subfields of the XMM-Large Scale Structure Survey field \citep[XMM-LSS;][]{Pierre04}.
The XMM subfields are the Subaru/XMM-Newton DEEP Survey field \citep[XMM-SXDS;][]{Furusawa08} and the Canada-France-Hawaii 
Telescope Legacy 
Survey (CFHTLS) field (XMM-CFHTLS).
These two fields are adjacent but are treated separately in our analysis 
as they were targeted by PRIMUS using different photometric catalogs \citep{Coil11}.
Full details of the survey design, targeting, and data summary can be found in \citet{Coil11}, while details of data reduction, redshift 
fitting, precision, 
and survey completeness are available in \citet{Cool13}.

Here we use the PRIMUS fields that have deep multi-wavelength ultraviolet (UV) imaging from the Galaxy Evolution Explorer 
\citep[GALEX;][]{Martin05}, 
mid-infrared imaging from the Spitzer Space Telescope \citep{Werner04} Infrared Array Camera \citep[IRAC;][]{Fazio04}, and optical and 
near-IR imaging 
from various ground-based surveys.
These include the CDFS, COSMOS, ES1, XMM-CFHTLS, and XMM-SXDS fields, covering $\sim5.5~\degsq$ on the sky.

\begin{figure*}
  \centering
  \includegraphics[width=0.8\textwidth,natwidth=600,trim={1.6in 0.4in 1.6in 0.6in},clip]{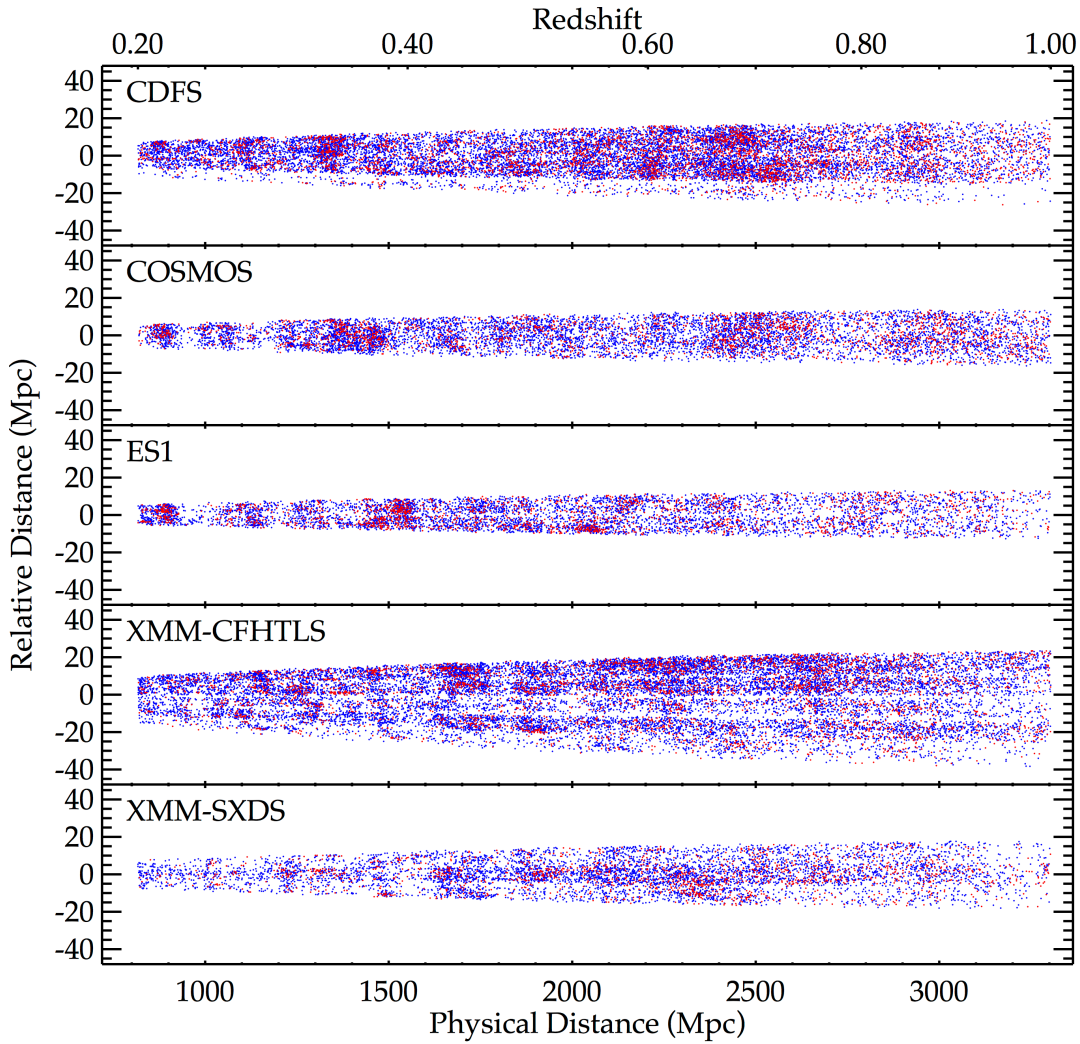}
  \caption{Redshift space distributions of PRIMUS galaxies as a function of physical distance along the line-of-sight and physical distance in the right ascension (RA) direction, relative to 
the median RA of the field.
Only galaxies with robust redshifts ${(Q \ge 3)}$ are shown.
Star-forming galaxies are shown in blue and quiescent galaxies in red (see~\S\ref{sec:SFQ}). Large-scale differences in the observed 
density of galaxies, 
for example, as a function of RA, reflect the number of slitmasks and targeting density.
}
  \label{fig:cone_diagrams}
\end{figure*}

\subsection{Full Sample and Targeting Weights}\label{sec:targ_weight}
 
Objects in PRIMUS are classified as galaxies, stars, or broad-line AGN by fitting the low-resolution spectra and multi-wavelength photometry 
for each source with an empirical library of templates.
The best-fit template defines both the redshift and the type of the source.  
We exclude AGN from this study and keep only those objects defined as galaxies with robust redshifts {($Q\ge 3$)} in the redshift range ${0.2<z<1.0}$.
We also only keep galaxies with well-defined targeting weights (these are termed ``primary'' galaxies in \citet{Coil11}; we do not use that naming here, to avoid 
confusion with our isolated primary sample defined below in \S\ref{sec:IPsample}).
These galaxies have a well-understood spatial and targeting selection function, defined by both a density-dependent weight and a magnitude-dependent sparse-sampling 
weight. In combination with a third, post-targeting weight that accounts for redshift incompleteness (see below) these weights allow a statistically complete galaxy sample to be recovered, which is suitable for analysis on two-point statistics, such as performed here.

PRIMUS targeting weights are described in detail in \citet{Coil11} and \citet{Cool13}.
Briefly, density-dependent weights account for sources that PRIMUS could not target in dense survey regions, as galaxies are sufficiently clustered in the plane of
the sky to the PRIMUS flux limit that even two slitmasks per pointing could not target every galaxy below the magnitude limit in each field (as spectra would overlap on the detector). 
Sparse-sampling weights are magnitude-dependent and ensure that the PRIMUS target catalog is not dominated by the faintest objects within the survey flux limit. Sparse-sampling weights were used to randomly select roughly a third 
of galaxies in the faintest 0.5~mag interval above the primary sample targeting limit.

\citet{Skibba2014} measured galaxy clustering in PRIMUS and tested the recoverability of two-point statistics with mock galaxy catalogs covering the PRIMUS survey volume.
They applied the same process used to select the PRIMUS target sample and calculate  density-dependent and sparse-sampling weights to a mock catalog, and generated a weighted mock sample. 
\citet{Skibba2014} then compared the correlation function of all galaxies in the mock catalog to that of the weighted mock sample and found no systematic difference between the two.  Thus when PRIMUS targeting weights are applied target selection does not impact the results presented in \S\ref{sec:results} below.

A third, post-targeting weight \citep[described in detail in][]{Cool13} accounts for the fact that not all PRIMUS spectra yielded reliable {($Q \ge 3$)} redshifts.
As shown in \S7 of \citet{Cool13}, the PRIMUS redshift success rate is primarily a function of $i$-band magnitude and does \emph{not} depend strongly on galaxy color. 
Taken together, the three weights described above allow for the recovery of a statistically complete galaxy sample from the targeted sources with reliable redshifts.

The full sample used here includes 60,071 galaxies in the five fields discussed above with robust redshifts between $0.2<z<1.0$ and well-understood selection weights.
Below we test the sensitivity of our results to these targeting and completeness weights.

\subsection{Stellar Mass and SFR Estimates}\label{sec:SFR}
 
Stellar masses and star formation rates (SFRs) of PRIMUS galaxies are obtained with SED fitting, a widely adopted method for estimating the physical properties of galaxies.
A complete description of the SED fitting process using \iSEDfit can be found in \citet{Moustakas13}, but we summarize the relevant points here.

\iSEDfit is a suite of routines written in the \IDL programming language that uses galaxy redshifts and photometry to compute the statistical likelihood of a large ensemble of model SEDs for each galaxy.
Model SEDs are generated using population synthesis models, and span a wide range of observed colors and physical properties (age, metallicity, star formation history, dust content, etc.).
\iSEDfit uses a Monte Carlo technique to randomly select values of model parameters from user-defined parameter distributions and compute a posterior probability distribution function (PDF).
PDFs of stellar mass and SFR are found by marginalizing over all other parameters, and the median value of the marginalized PDF is taken as the best estimate of the stellar mass or SFR of each galaxy.

To test how the uncertainties on the stellar mass and SFR estimates described above affect our classification of galaxies as either star-forming or quiescent, we randomly sampled individual stellar masses and SFRs for each galaxy in the full sample 100 times from normal distributions with widths equal to the stellar mass or SFR error for that galaxy.  Over 100 trials there is an average change in the star-forming fraction of $<1\%$.  Below in \S\ref{sec:signal} we discuss how this small difference may affect our conformity results.

\subsection{Identifying Star-forming and Quiescent Galaxies}\label{sec:SFQ}

\begin{figure}
  \centering
  \includegraphics[width=\linewidth,natwidth=600,trim={0.1in 0.1in 0.2in 0.6in},clip]{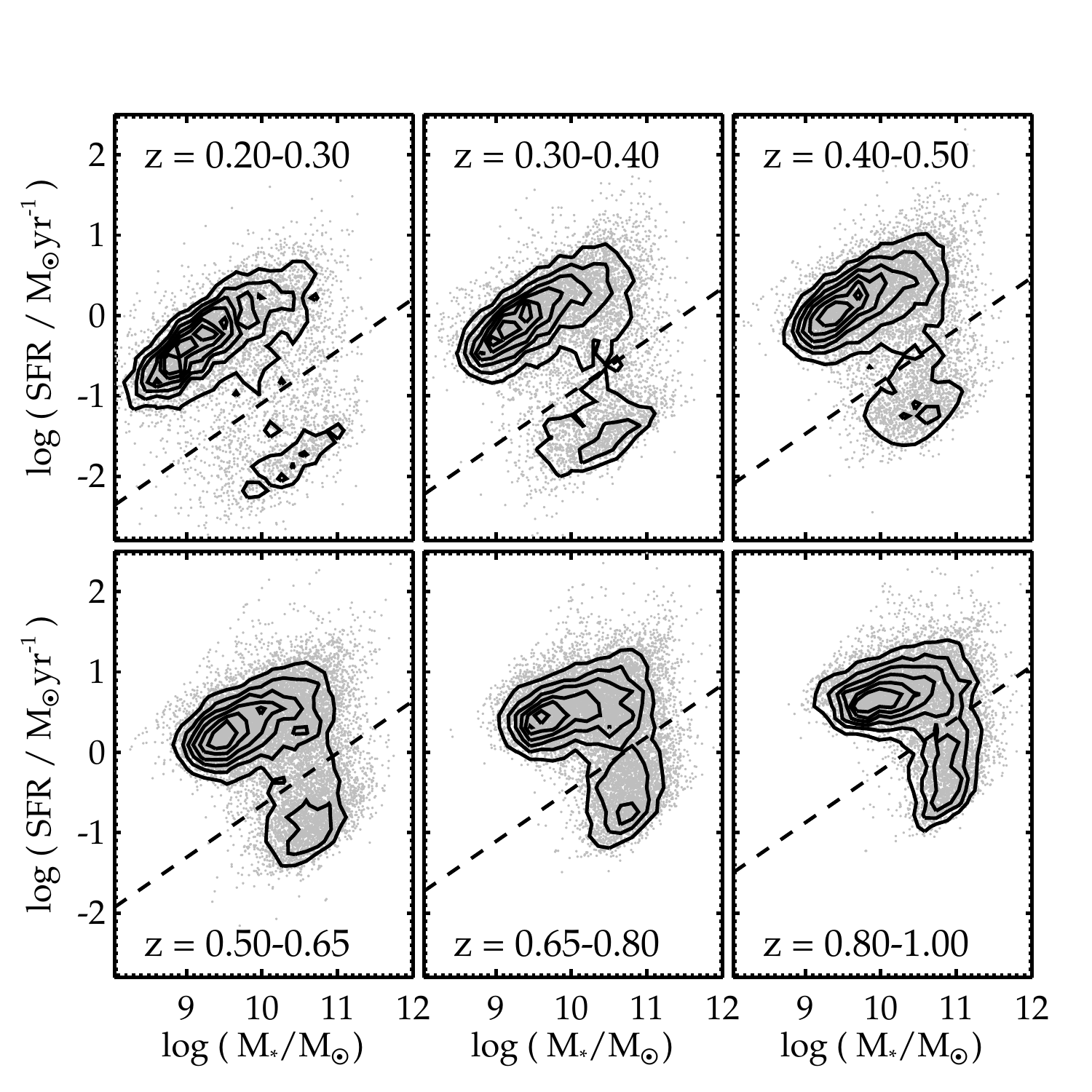}
  \caption{Star formation rate (SFR) versus stellar mass for PRIMUS galaxies in six redshift bins from ${z=0.2}$--1.
Galaxies in our sample are classified as star-forming or quiescent according to whether they lie above or below the dashed line, respectively.
This line runs parallel to the star-forming main sequence, traces the minimum in the galaxy SFR bimodality, and evolves with redshift according to Equation~\ref{eq:SFR}.
}
  \label{fig:SFR_vs_mass}
\end{figure}

We divide our sample into star-forming and quiescent galaxies based on each galaxy's position in the SFR\textendash stellar mass plane. 
Figure~\ref{fig:SFR_vs_mass} shows SFR versus stellar mass in six redshift bins from ${z=0.2}$--1 for the PRIMUS galaxy sample.
The dashed line (Eq.~\ref{eq:SFR}) in each panel traces the minimum of the bimodal galaxy distribution in that bin and is given by the following linear relation:
\begin{equation}
\log\,({\rm SFR}) = -1.29 + 0.65\,\log\,(\mass - 10) + 1.33\,(z - 0.1)
\label{eq:SFR}
\end{equation}

\noindent where SFR has units of $\sfrunit$ and $\mass$ has units of $\msun$.
The slope of this line is defined by the slope of the star-forming main sequence \citep{Noeske07} as measured in the PRIMUS dataset using \iSEDfit SFR and stellar mass estimates. 
Each galaxy is classified as star-forming or quiescent based on whether it lies above or below the cut defined by Equation~\ref{eq:SFR}, evaluated at the redshift of the galaxy.

\subsection{Isolated Primary Sample}\label{sec:IPsample}

In order to measure galactic conformity we must first identify isolated galaxies around which to search for the signal.
We follow K13, who selected in SDSS a volume-limited sample of galaxies with $\logM>9.25$ and $0.017<z<0.03$.  They then defined ``central" galaxies of stellar mass $\mstar$ as those in their sample with no other galaxies with stellar mass greater than $\mstar/2$ within a projected radius of 500~kpc and with a velocity difference less than 500~\kms.  Any galaxy in our full sample (defined above) is considered an isolated primary (IP) if there are no other galaxies
(i) within a projected physical distance of 500~kpc from the IP candidate,
(ii) within ${\pm 2.0\,\sigma_{z}\,(1 + z_{\rm IP})}$ in redshift space from the IP candidate (this includes as many true neighbors as possible while simultaneously minimizing interlopers and integrates over peculiar velocities), and
(iii) with stellar mass greater than half the stellar mass of the IP candidate.
Additionally, IPs can be neighbors (see \S\ref{sec:LTfraction}) of other IPs, and all galaxies can be a neighbor of multiple IPs. 

It is possible for galaxies near the edge of the survey area to be incorrectly classified as isolated if they have a sufficiently massive neighbor within a projected physical distance of 500~kpc that lies outside the survey area.
This could lead to contamination of our IP samples.
To test for this potential effect we visually inspected the distribution of IPs near the survey edges and concluded that false detections near edges do not significantly impact our IP sample, in that the spatial density of IPs does not rise substantially at the survey edges. 

\subsubsection{Stellar Mass Completeness Limits}\label{sec:mass_limit}

Because PRIMUS is a flux-limited survey targeted in the $i$ band, galaxies with higher SFRs (i.e.~bluer galaxies) can be more easily detected at lower stellar mass than galaxies with lower SFR (i.e.~redder galaxies).
This introduces a bias towards star-forming galaxies in the PRIMUS sample at lower stellar masses.
To account for this bias we define a stellar mass limit above which at least 95\% of all galaxies can be detected, regardless of SFR.
This stellar mass completeness limit is a function of redshift, galaxy type (star-forming or quiescent), and also varies slightly between fields (due to the different photometry used for targeting in each field).
Details of the calculation of PRIMUS mass completeness limits can be found in \citet{Moustakas13}.
Briefly, we compute the stellar mass each galaxy would have if its apparent magnitude were equal to the survey magnitude limit, $\mlim$.  We then construct the cumulative distribution of $\mlim$ for the 15\% faintest galaxies in redshift bins of width ${\Delta z=0.04}$, and calculate the minimum stellar mass that includes 95\% of the objects.  The limiting stellar mass versus redshift is fit with a separate quadratic polynomial for all, star-forming, and quiescent galaxies, and the fit is evaluated at the center of each redshift interval \citep[see][]{Moustakas13}.

In addition to the isolation criteria described above, all IPs must have stellar masses above the stellar mass completeness threshold specific to the IP's field, redshift, and type (star-forming or quiescent; see Table~\ref{table:mass_comp_limit}).  When identifying IPs, for each field and each redshift range in Table~\ref{table:mass_comp_limit} we eliminated any star-forming (quiescent) galaxy with stellar mass below the limiting value for star-forming (quiescent) galaxies in that field and redshift range.

Of the 60,071 galaxies in the full sample, 14,888 star-forming and 6,847 quiescent galaxies meet the isolation and stellar mass completeness criteria to be IPs.

\input{tab01_mass_comp_lim}

\subsubsection{Matching Stellar Mass and Redshift}\label{sec:IPsample_matching}

While our star-forming and quiescent IP populations are statistically complete (after applying the targeting and completeness weights), even above the stellar mass completeness limits the median stellar masses and redshifts of the two populations differ, as the stellar mass functions of star-forming and quiescent galaxies are different.

\begin{figure}
  \epsscale{1.1}
  \epstrim{0.1in 0.1in 0.5in 0.8in}
  \plotone{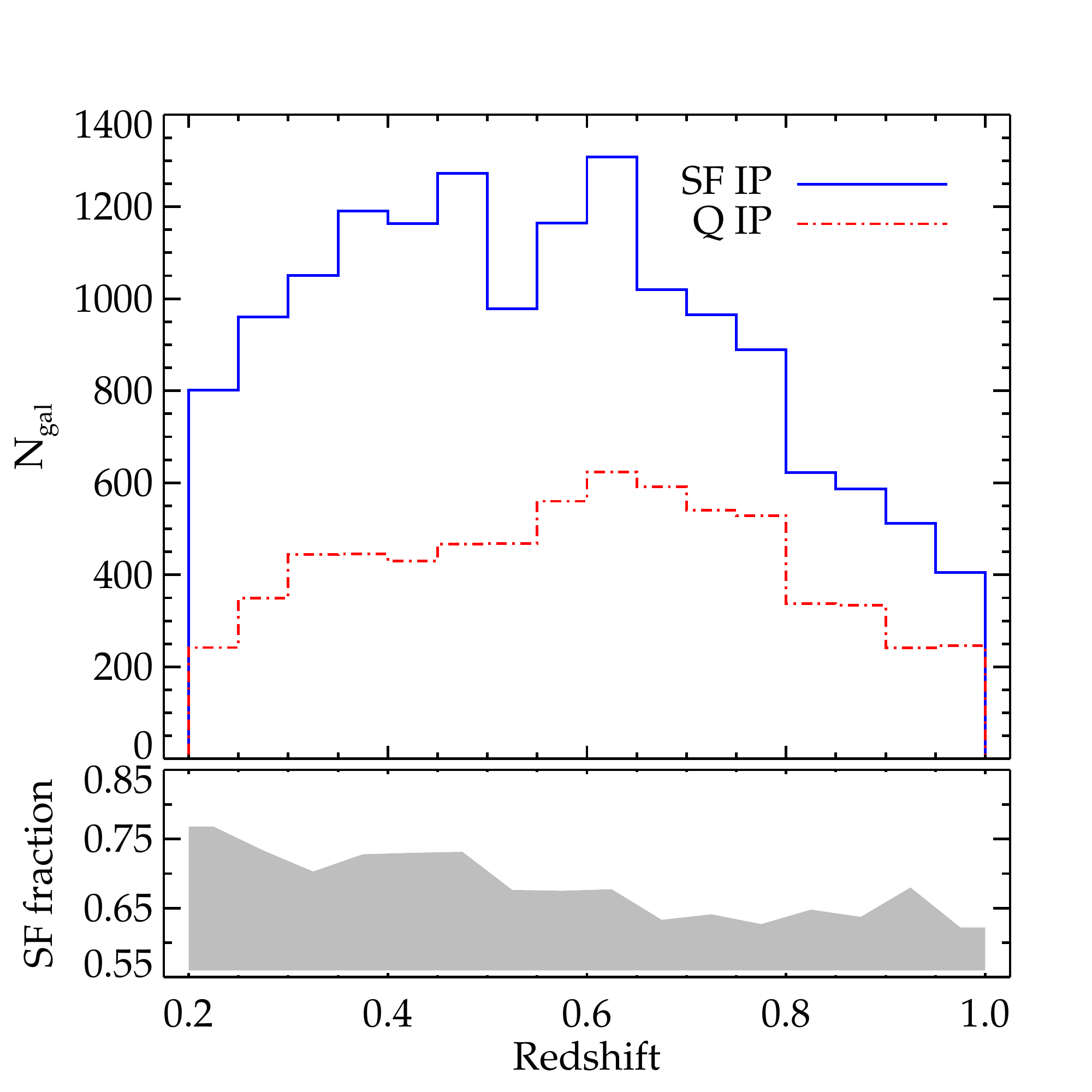}
  \caption{Top panel: Redshift histograms of all star-forming (solid blue line) and quiescent (dash-dot red line) IPs.
Bottom panel: Star-forming fraction of all PRIMUS galaxies in the full sample as a function of redshift. 
}
  \label{fig:IPhist_latefrac_vs_z}
\end{figure}

Figure~\ref{fig:IPhist_latefrac_vs_z} shows the redshift distributions of all star-forming (solid blue line) and quiescent (dashed red line) IPs, and the star-forming fraction of all PRIMUS galaxies in the full sample as a function of redshift.
Our star-forming and quiescent IP populations have median stellar masses of $\logM=10.44$ and 10.86, respectively, and median redshifts of $z=0.55$ and 0.60.

Several recent studies (as well as this work) caution that systematic errors can in some cases create an artificial conformity signal.
For example, \citet{Campbell15} found that while group-finding algorithms do a good job of recovering one-halo galactic conformity in mock catalogs, they also have a tendency to introduce a weak conformity signal when none is present.

As discussed below, to compare the star-forming fraction of neighbors around star-forming and quiescent IP galaxies we require the star-forming and quiescent IP samples to have the same stellar mass and redshift distributions.
To obtain these ``matched'' IP samples we first apply to our IP samples an {\it upper} stellar mass cut derived from the PRIMUS stellar mass function \citep[SMF, denoted $\Phi$; see][]{Moustakas13} for star-forming galaxies.
This upper cut is required as there are fewer star-forming galaxies at high stellar mass ($\logM>11$) than quiescent galaxies.
Therefore the high-mass end of the star-forming galaxy SMF defines the upper stellar mass limit of our matched IP samples.  
Specifically, we eliminate all IPs (both star-forming and quiescent) with stellar masses greater than the stellar mass at which 
${\log\,(\Phi \,/\, 10^{-4}\,\text{Mpc}^{-3}\,\text{dex}^{-1}) \le -3.7}$, interpolated at the redshift of each galaxy.
These upper mass limits are listed in Table~\ref{table:SMFlimit}.

\input{tab02_SMF_lim}

\begin{figure}
  \epsscale{1.1}
  \epstrim{0.4in 0.2in 0.2in 0.4in}
  \plotone{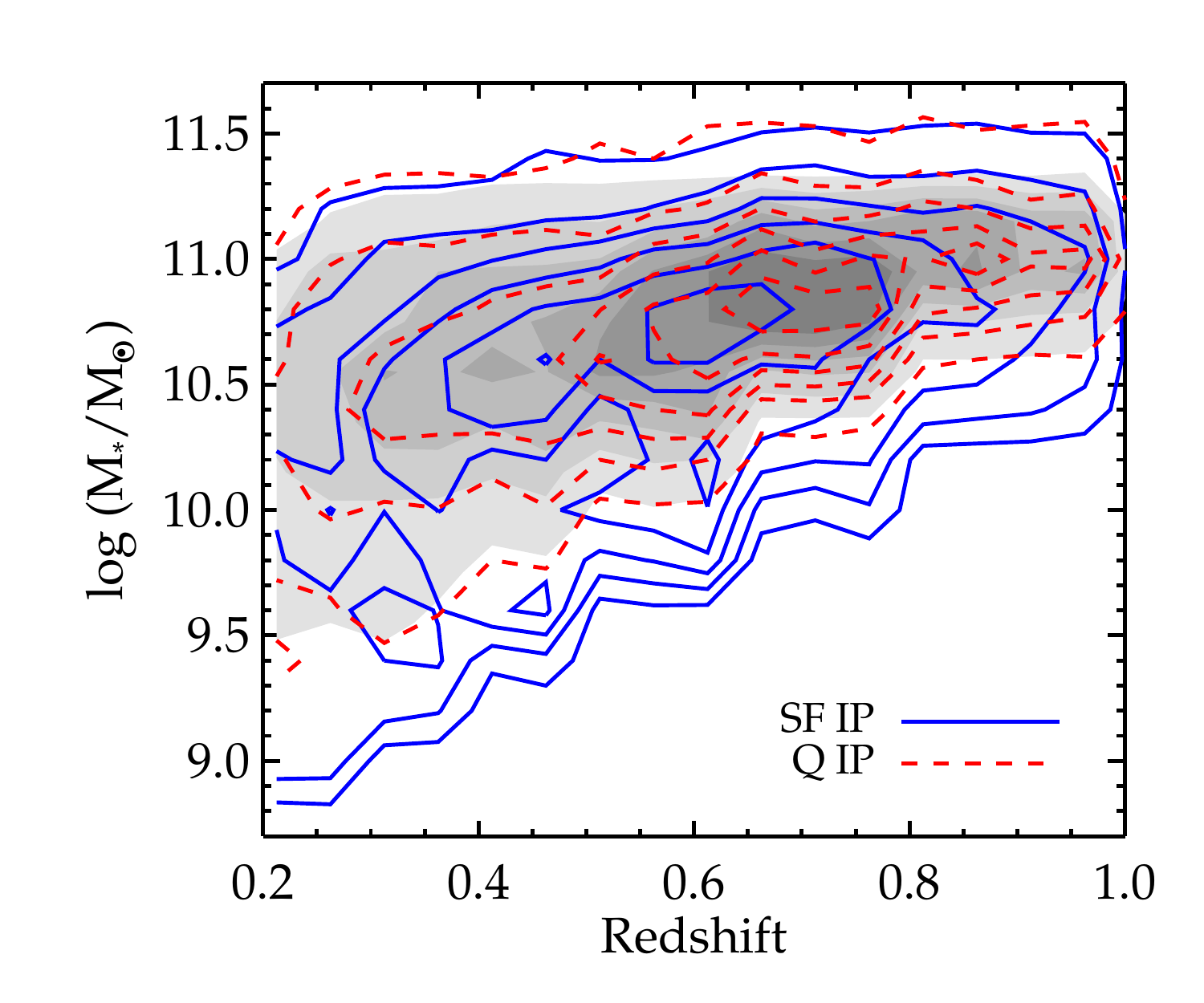}
  \caption{Stellar mass and redshift distribution for all star-forming (blue solid contours) and quiescent (red dashed contours) IPs in the full sample. 
Gray shaded contours show the matched sample, in which star-forming and quiescent IPs have the same stellar mass and redshift distributions.
} 
  \label{fig:IPsample_matched}
\end{figure}

We then create a two-dimensional histogram of the stellar mass and redshift distribution of the remaining quiescent IP population, in bins of 0.2~dex in stellar mass and 0.05 in redshift.
For each of our five fields, in each bin we randomly select with replacement the same number of star-forming as there are quiescent IPs.
This selection is done separately in each field to account for field-to-field variations in the stellar mass and redshift distributions of the IP populations.
Our final matched IP sample (hereafter ``matched sample'') contains 6,197 unique quiescent and 4,185 unique star-forming IPs.
Each star-forming IP is assigned a weight equal to the number of times it was randomly selected while matching the distribution of the quiescent IP sample.
The sum of all star-forming IP weights therefore equals the total number of unique quiescent IPs.
Figure~\ref{fig:IPsample_matched} shows the stellar mass and redshift distributions of all star-forming and quiescent IPs in the full sample, as well as the stellar mass and redshift distributions of the matched sample.

%% file: tab01_mass_comp_lim.tex
\setlength{\tabcolsep}{0.02in}
\begin{deluxetable}{cccccc}
\tablecaption{Stellar mass completeness limits for star-forming and quiescent galaxies as a function of redshift.  At least 95\% of all galaxies above these stellar mass limits can be detected regardless of SFR.
\label{table:mass_comp_limit}}
\tablewidth{0pt}
\tablehead{
\colhead{}	&									
\colhead{\scriptsize CDFS}	&
\colhead{\scriptsize COSMOS}	&
\colhead{\scriptsize ES1}	&
\colhead{\scriptsize XMM-CFHTLS} &
\colhead{\scriptsize XMM-SXDS} \\
\cline{2-6} \\										
\colhead{Redshift Range} &
\multicolumn{5}{c}{$\log\,(\mlim/\msun)$}
}
\startdata
\multicolumn{6}{c}{Star-Forming} \\
\cline{1-6} \\[-1ex]
$0.20-0.30$ &	9.60	&	8.68	&	9.58	&	8.80	&	8.79	\\
$0.30-0.40$ &	9.92	&	9.05	&	9.94	&	9.06	&	9.13	\\
$0.40-0.50$ &	10.19	&	9.38	&	10.25	&	9.30	&	9.44	\\
$0.50-0.65$ &	10.44	&	9.75	&	10.59	&	9.58	&	9.77	\\
$0.65-0.80$ &	10.63	&	10.12	&	10.90	&	9.89	&	10.10	\\
$0.80-1.00$ &	10.69	&	10.46	&	11.14	&	10.21	&	10.38	\\
\cutinhead{Quiescent}										
$0.20-0.30$ &	9.65	&	9.23	&	9.80	&	9.17	&	9.35	\\
$0.30-0.40$ &	9.92	&	9.58	&	10.06	&	9.52	&	9.61	\\
$0.40-0.50$ &	10.17	&	9.89	&	10.30	&	9.85	&	9.85	\\
$0.50-0.65$ &	10.44	&	10.22	&	10.55	&	10.22	&	10.13	\\
$0.65-0.80$ &	10.71	&	10.52	&	10.79	&	10.60	&	10.43	\\
$0.80-1.00$ &	10.96	&	10.75	&	10.99	&	10.96	&	10.73	\\
\enddata
\end{deluxetable}

%% file: tab02_SMF_lim.tex
\setlength{\tabcolsep}{0.1in}
\begin{deluxetable}{cc}
\tablewidth{0pc}
\tablecolumns{2}
\tablecaption{Upper stellar mass limit for galaxies in the matched IP sample.
\label{table:SMFlimit}
}
\tablehead{
\colhead{Redshift Range} & \colhead{$\log\,(\mmax / \msun)$} }
\label{table:SMFlimit}
\startdata
$0.20-0.30$ & 11.154 \\
$0.30-0.40$ & 11.208 \\
$0.40-0.50$ & 11.255 \\
$0.50-0.65$ & 11.241 \\
$0.65-0.80$ & 11.308 \\
$0.80-1.00$ & 11.324 \\
\enddata
\end{deluxetable}

%% file: results.tex
\section{Results}\label{sec:results}

In this section we discuss the importance of matching star-forming and quiescent IP 
samples in both stellar mass and redshift, and we present the one- and two-halo 
conformity signal in the matched PRIMUS sample.  We discuss the effects of 
cosmic variance on measures of conformity and the need for jackknife
errors at intermediate redshifts, and we investigate the redshift and stellar mass 
dependence of conformity within the PRIMUS sample.

\subsection{The Effects of Matching Redshift and Stellar Mass on the Conformity Signal}\label{sec:LTfraction}

As discussed above, galactic conformity is the observed tendency of neighbor 
galaxies to have the same star-formation type (star-forming or quiescent) 
as their associated IP galaxy.
One-halo conformity refers to conformity between an IP and the neighbors within the 
same dark matter halo (i.e.~within $\sim0.5$~Mpc of the IP),
while two-halo conformity refers to conformity between an IP and neighbors in other 
adjacent halos (i.e.~at distances greater than $\sim0.5$~Mpc from the IP).

We therefore want to measure how the fraction of neighbors that are star-forming differs 
between star-forming and quiescent IP hosts as a function of projected radius from 
the IP.
To do this, for each IP in our matched sample we count all neighbors within 
concentric cylindrical shells of length ${2\times2\,\sigma_{z}(1+z_{\text{IP}})}$
and cross-sectional area
${\pi[(\Rproj+d\Rproj)^2-\Rproj^2]}$, where $\Rproj$ is the 2D projected radius from the IP in (physical) Mpc, and $d\Rproj$ is the shell width in Mpc.
The star-forming fraction of neighbors of star-forming IPs in a cylindrical shell 
at projected radius $\Rproj$ to ${(\Rproj+d\Rproj)}$, $f^{\textrm{SF-IP}}_{\textrm{SF}}
(\Rproj)$ is defined to be the sum of the targeting weights (see~\S\ref{sec:targ_weight}) of the star-forming neighbors of star-forming IPs in the shell, 
divided by the sum of the
targeting weights of \emph{all} neighbors of star-forming IPs in the shell:
\begin{equation}
        f^{\textrm{SF-IP}}_{\textrm{SF}}(\Rproj) = \frac
        {\displaystyle \sum_{i=1}^{N_{\textrm{SF-IP}}} \sum_{j=1}^{N_{\textrm{SF},i}} w_{j} }
        {\displaystyle \sum_{i=1}^{N_{\textrm{SF-IP}}} \sum_{k=1}^{N_{\textrm{tot},i}} w_{k} },
\label{eq:SFfrac}
\end{equation}
and likewise for quiescent IPs.
$N_{\textrm{SF-IP}}$ is the total number of star-forming IPs, $N_{\textrm{SF},i}$ is the number of star-forming neighbors of IP $i$ in the shell, $N_{\textrm{tot},i}$ is the
total number of neighbors of IP $i$ in the shell, and $w_{j}$ and $w_{k}$ are PRIMUS targeting weights of the neighbors.
We are therefore essentially computing star-forming neighbor fractions for star-forming 
and quiescent IPs by stacking the neighbors of all IPs of each type.

\begin{figure*}
  \epsscale{0.9}
  \epstrim{0.1in 0.3in 0.4in 0.8in}
  \plotone{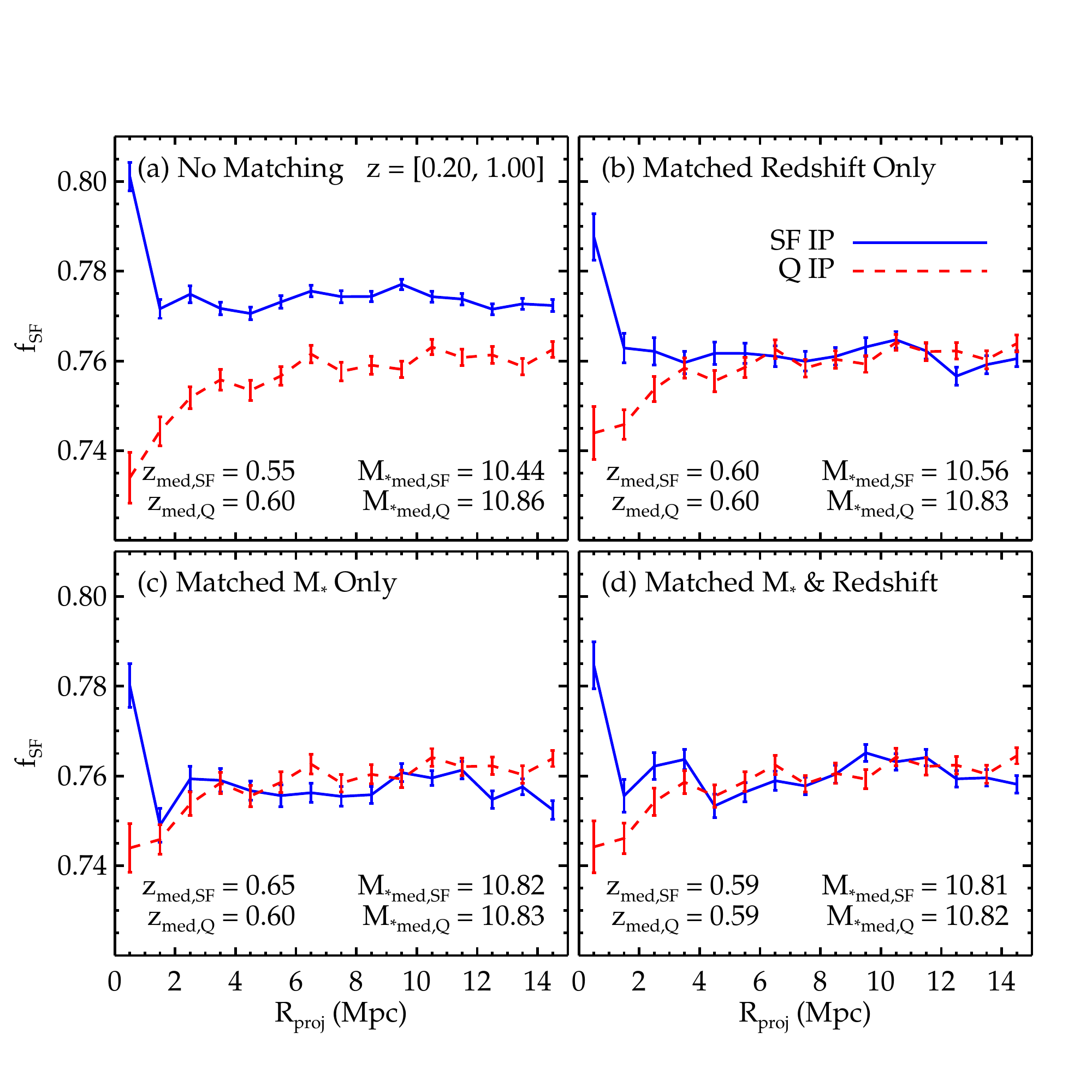}
  \caption{
The fraction of star-forming neighbor galaxies around star-forming 
and quiescent IPs, to a projected distance of ${\Rproj<15}$~Mpc, 
 for four different IP samples: 
(a)~all IP candidates above the \citet{Moustakas13} mass completeness limit (\S\ref{sec:mass_limit});
(b)~IP candidates that also have the same redshift distribution for the star-forming
and quiescent IPs; 
(c)~IP candidates that have the same stellar mass distribution;
(d)~IPs that have both matched stellar mass and redshift distributions.
The median redshift and stellar mass of each IP sample are shown in each panel.
Errors are computed by bootstrap resampling as described in the text.
}
  \label{fig:IPsample_compare}
\end{figure*}

The importance of matching both the stellar mass and redshift distributions of our 
IP sample is clearly illustrated in Figure~\ref{fig:IPsample_compare}, which shows 
how the star-forming fractions of neighbors around star-forming and quiescent IPs 
differ when different IP samples are used.
Figure~\ref{fig:IPsample_compare} shows the fraction of neighbors of star-forming 
and quiescent IPs that are star-forming as a function of projected radius from the 
IP in 1~Mpc annuli out to 15~Mpc for four different IP samples.

In panel (a) all IP candidates above the \citet{Moustakas13} mass completeness limit 
(\S\ref{sec:mass_limit}) are included.
Here the median stellar mass of the quiescent IP population is 0.42~dex greater 
than that of the star-forming IP population, and the median redshift is greater 
by 0.05.
This difference in the stellar mass distribution in particular means that 
star-forming IPs are preferentially located at lower redshift, where the star-forming fraction of \emph{all} PRIMUS galaxies (the ``full'' sample; see~\S\ref{sec:targ_weight}) is larger than at higher redshifts.  The star-forming fraction of the full sample declines steadily from $\sim0.80$ at $z\sim0.2$ to $\sim0.73$ at $z\sim1.0$, causing us to overestimate the star-forming neighbor fraction for star-forming IPs at all projected radii.
The result is a relatively fixed offset between the solid and dashed lines in the 
upper left panel of Figure~\ref{fig:IPsample_compare} that persists to the largest 
projected radii we measure with PRIMUS, mimicking a conformity signal.  We therefore
measure a ``false'' conformity signal in this sample.

In panel (b) we select star-forming and quiescent IP samples with matched redshift 
distributions using the method described in \S\ref{sec:IPsample_matching}.  This 
eliminates the large-scale offset, but there is still a 0.3~dex difference in the 
median stellar masses of the IP samples.
Since star-forming fraction depends on stellar mass, this is not ideal. 

In panel (c) we select star-forming and quiescent IP samples with matched stellar mass distributions; this results in a star-forming IP sample with a higher median 
redshift than that of the quiescent IP sample (by 0.05).
In this case the systematic bias mimics the opposite of a conformity signal:~the solid line moves closer to the dashed line at all projected radii, actually dropping below it at $\gtrsim5$~Mpc.

Finally, panel (d) shows results for our matched stellar mass and matched redshift IP sample.
Failing to control for differences in stellar mass and/or redshift distributions 
can introduce bias into the relative star-forming neighbor fractions of star-forming and quiescent IPs.
Only by matching both the stellar mass and redshift distributions of our star-forming and quiescent IP samples do we eliminate systematic biases in star-forming neighbor fraction measurements that could masquerade as a conformity signal.

For the remainder of this paper, the IP samples matched in both stellar mass and redshift are referred to as the ``matched'' sample.

\subsection{One- and Two-Halo Conformity Signal in Matched Sample}\label{sec:signal}

\begin{figure*}
  \epsscale{0.75}
  \epstrim{0.4in 0.1in 0.3in 0.4in}
  \plotone{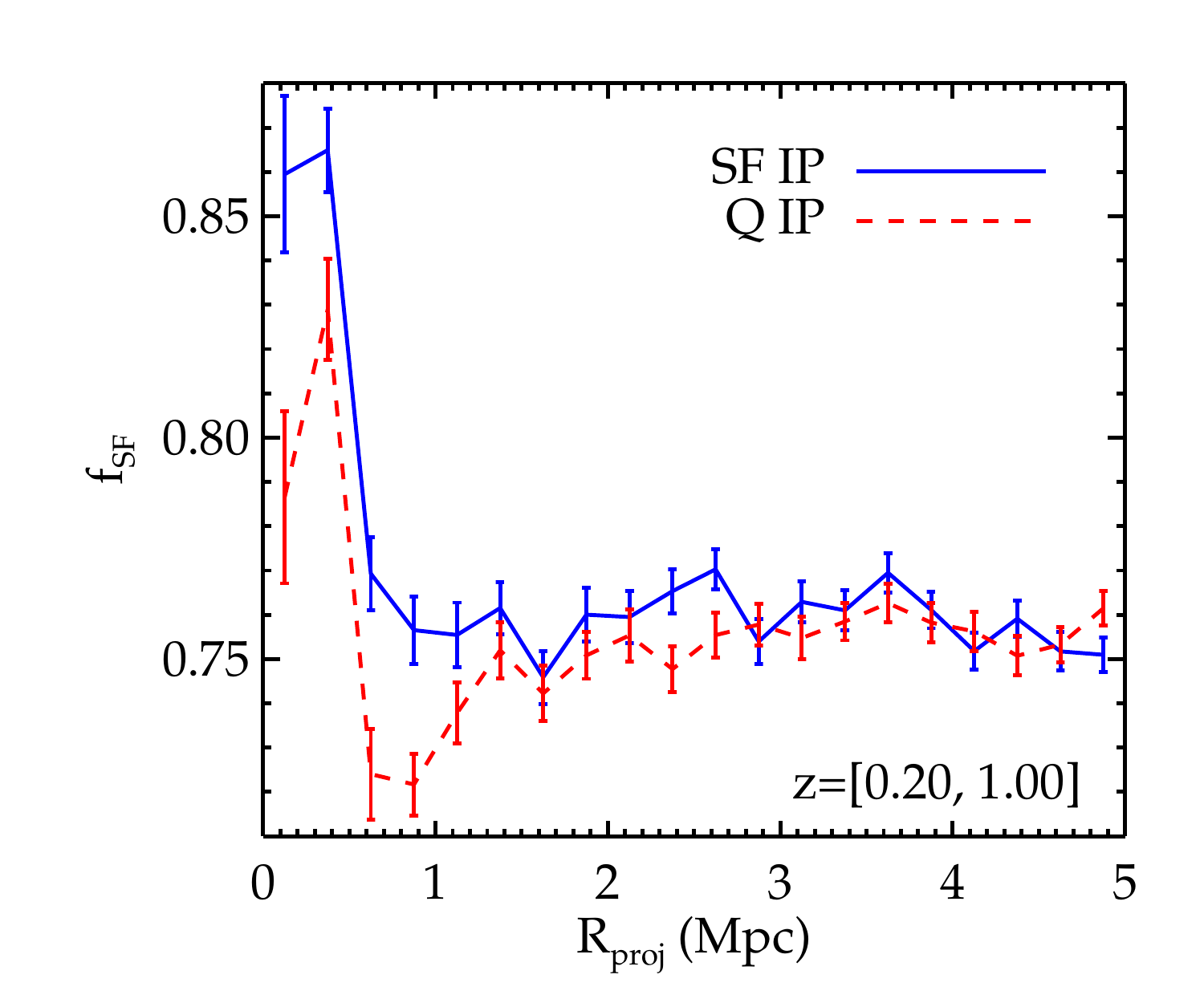}
  \caption{
The fraction of star-forming neighbor galaxies around star-forming 
and quiescent IPs to a projected distance of ${\Rproj<5}$~Mpc 
for IP samples matched in both stellar mass and redshift.  Here we use
finer radial bins (${d\Rproj=0.25}$~Mpc) for all star-forming (blue solid line) and quiescent (red dashed line) IPs in the matched sample.
Errors are computed by bootstrap resampling.
}
  \label{fig:latefrac_matched}
\end{figure*}

Stacked star-forming neighbor fractions for the matched sample of star-forming and quiescent IPs are shown in Figure~\ref{fig:latefrac_matched}, here using finer radial bins.  The errors here and above are estimated by bootstrap resampling, where for each radial bin we randomly select with replacement 90 percent of all star-forming or quiescent IPs 200 times, and compute $\flate$ for each of the 200 samples.  The bootstrap error is the standard deviation of the $\flate$ distribution.  Below in \S\ref{sec:errors} we discuss the merits of estimating error with jackknife versus bootstrap resampling.

In Figure~\ref{fig:latefrac_matched} the one-halo component of the conformity signal is clearly visible as the 4--7\% difference between $\flate$ for star-forming and quiescent IPs at $\Rproj < 1$~Mpc.  Within this range $\flate$ for both IP types is greatest at $\Rproj<0.5$~Mpc:~$\sim86\%$ for star-forming and $\sim80\%$ for quiescent IPs.  At $\Rproj=0.5$~Mpc $\flate$ for both IP types drops sharply by at least 8\% to $\sim76\%$ for star-forming and $\sim72\%$ for quiescent IPs.

This break at 500~kpc is an artifact of the isolation criteria used to identify isolated primaries 
(see \S\ref{sec:IPsample}), and the fact that the fraction of all galaxies in our full sample that are
star-forming is a decreasing function of stellar mass.
Because we require IP galaxies (regardless of type) to have no other galaxies more massive than half the
stellar mass of the IP within 500 projected kpc, the median stellar mass of galaxies within 500~kpc will
automatically be lower than the median stellar mass of galaxies beyond this distance.
The star-forming fraction of neighboring galaxies within 500~kpc will therefore be greater than the star-forming
fraction of neighbors within ${0.5<\Rproj<5}$~Mpc.

To confirm that this feature of Figure~\ref{fig:latefrac_matched} is a direct result of our choice of a
projected radius of 500~kpc when identifying IPs, we also measured $\flate$ for redshift and stellar 
mass-matched samples of star-forming and quiescent IPs selected using 250 and 750~kpc as the projected 
radius for our isolation criteria.
As expected, when 250~kpc is used to identify IPs, the break in $\flate$ 
for both star-forming and quiescent IPs occurs at 250~kpc, and likewise for 750~kpc.
Because conformity is the \emph{difference} between $\flate$ for star-forming and quiescent IPs and does
not depend on the absolute star-forming neighbor fraction for either IP type, this break at 500~kpc does
not affect our result.

Over $\Rproj\simeq1$--1.5~Mpc $\flate$ for quiescent IPs increases to $\sim0.75$, while for star-forming IPs $\flate$ begins to level off at $\sim0.76$.  This $\sim1$\% difference between the two fractions is a two-halo conformity signal that persists to to roughly 3~Mpc.  At $3\lesssim\Rproj<5$~Mpc $\flate$ for both IP types is effectively constant and nearly equal, such that no conformity signal is present beyond $\Rproj\simeq3$~Mpc.

\begin{figure}
  \epsscale{1.1}
  \epstrim{0.5in 0.1in 0.3in 0.3in}
  \plotone{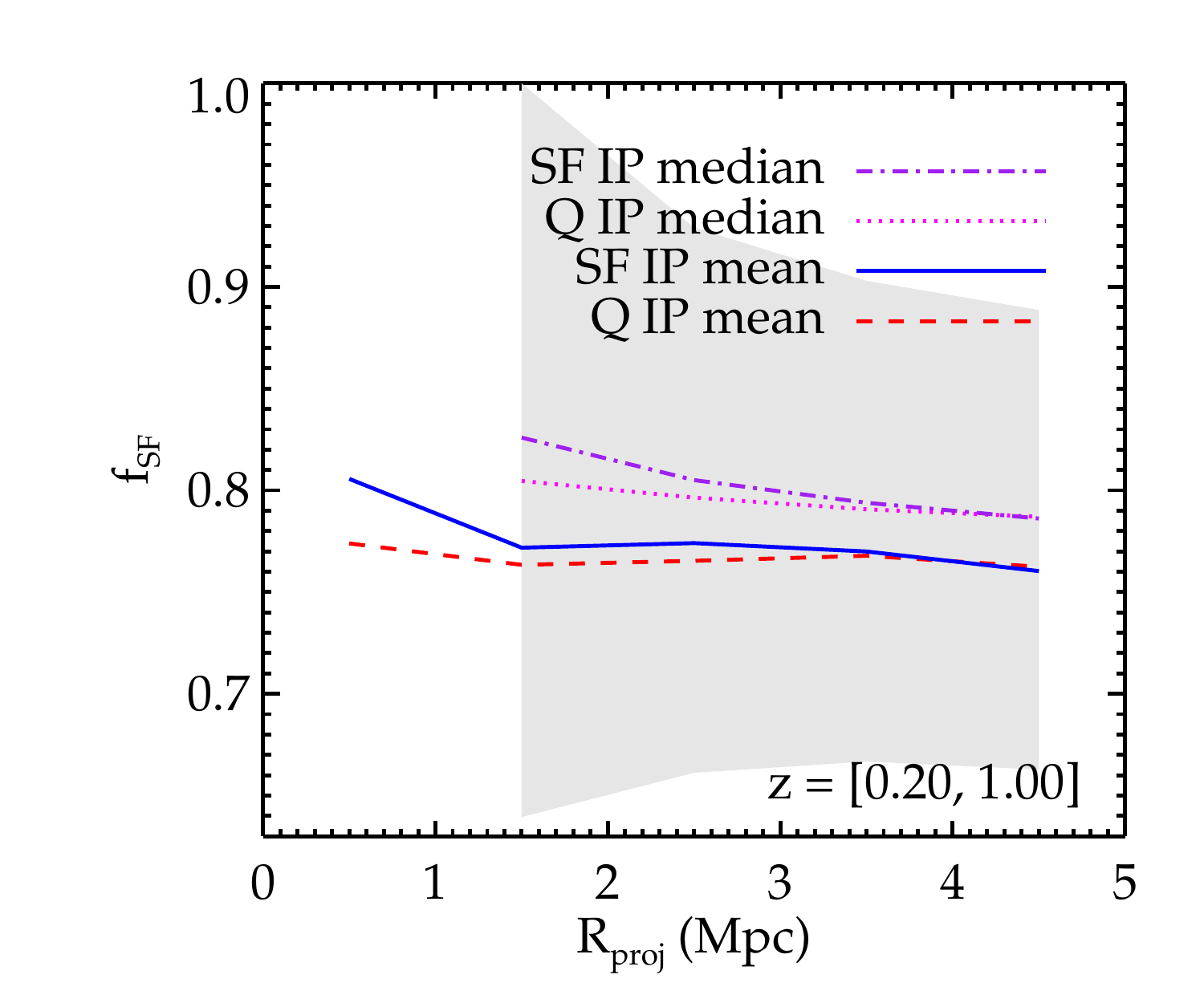}
  \caption{
Similar to Figure~\ref{fig:latefrac_matched}, except here $\flate$ is the median of 
the distribution of non-zero individual star-forming neighbor fractions for each IP type as a function of $\Rproj$.  This effectively gives equal weight to each IP, instead of upweighting the IPs with more neighbors, as shown in Figure 6.
Also shown is the mean of the non-zero star-forming neighbor fraction distributions of star-forming and quiescent IPs (purple dash-dot and magenta dotted lines),
and the interquartile range of the combined distribution for both IP types (gray shaded region).
}
  \label{fig:latefrac_quartiles}
\end{figure}

Within a particular radial bin (or shell around each IP), this stacking method 
weights IPs with more neighbors more heavily than those with fewer or no neighbors.
To assess whether this will bias our results, we recompute the star-forming neighbor fraction, 
now assigning equal weight to each IP by computing the star-forming fraction 
individually for each IP and then taking the median of the distribution of all 
non-zero fractions for both IP types, within each 1~Mpc radial bin.
The result is shown in Figure~\ref{fig:latefrac_quartiles}, which also shows the 
mean individual star-forming neighbor fraction for both IP types in each radial bin (again using
only non-zero fractions), and the interquartile range (25$^{\textrm{th}}$ to 75$^{\textrm{th}}$ percentile) of the combined $\flate$ distribution for both IP types.

Above $\Rproj=1$~Mpc the large spread in the interquartile range indicates that three quarters of IPs have a star-forming neighbor fraction of at least 65\%, while for one quarter of IPs the star-forming fraction is over 90\%.
Median and interquartile range values are not shown for ${\Rproj<1}$~Mpc because in that bin the median (and 75$^{\textrm{th}}$ percentile) value of $\flate$ for both IP types is 1.

In the $\Rproj<1$~Mpc bin the only meaningful measure of conformity is the mean values of the $\flate$ distributions, which vary from $\sim81\%$ for star-forming IPs to $\sim77\%$ for quiescent IPs.  The difference between the mean values of $\flate$ decreases on scales of 1--3~Mpc and disappears entirely on larger scales.

The difference between star-forming neighbor fractions for star-forming and quiescent IPs is comparable for equal weighting of IPs as shown here and when each IP is weighted proportionally to its number of neighbors, as shown in Figure~\ref{fig:latefrac_matched}.

\input{tab04_signal_matched}

We define the normalized conformity signal, 
$\signorm$, at a projected radius of $\Rproj$ as the difference between the star-forming neighbor fractions of star-forming and quiescent IPs, 
divided by the mean of these two fractions:
\begin{equation}
	\signorm(\Rproj) = \frac
	{ \left| f^{\textrm{SF-IP}}_{\textrm{SF}}-f^{\textrm{Q-IP}}_{\textrm{SF}} \right| }
	{ \left( f^{\textrm{SF-IP}}_{\textrm{SF}}+f^{\textrm{Q-IP}}_{\textrm{SF}} \right) \! /2}
\label{eq:signorm}
\end{equation}

\noindent We note that the choice to define $\signorm$ in terms of the star-forming instead of the quiescent neighbor fraction is arbitrary.
Were we to instead define $\signorm$ as the normalized difference in \emph{quiescent} neighbor fraction ${f_{\textrm{Q}} = 1 - f_{\textrm{SF}}}$ then Equation~\ref{eq:signorm} would be
\begin{equation}\nonumber
	\frac
	{ \left| f^{\textrm{SF-IP}}_{\textrm{Q}}-f^{\textrm{Q-IP}}_{\textrm{Q}} \right| }
	{ \left( f^{\textrm{SF-IP}}_{\textrm{Q}}+f^{\textrm{Q-IP}}_{\textrm{Q}} \right) \! /2} = 
	\frac
	{ \left| f^{\textrm{SF-IP}}_{\textrm{SF}}-f^{\textrm{Q-IP}}_{\textrm{SF}} \right| }
	{\left[ 1 - \left( f^{\textrm{SF-IP}}_{\textrm{SF}}+f^{\textrm{Q-IP}}_{\textrm{SF}} \right) \! /2 \right]}.
\end{equation}

\begin{figure}
  \epsscale{1.1}
  \epstrim{0.3in 0.1in 0.2in 0.3in}
  \plotone{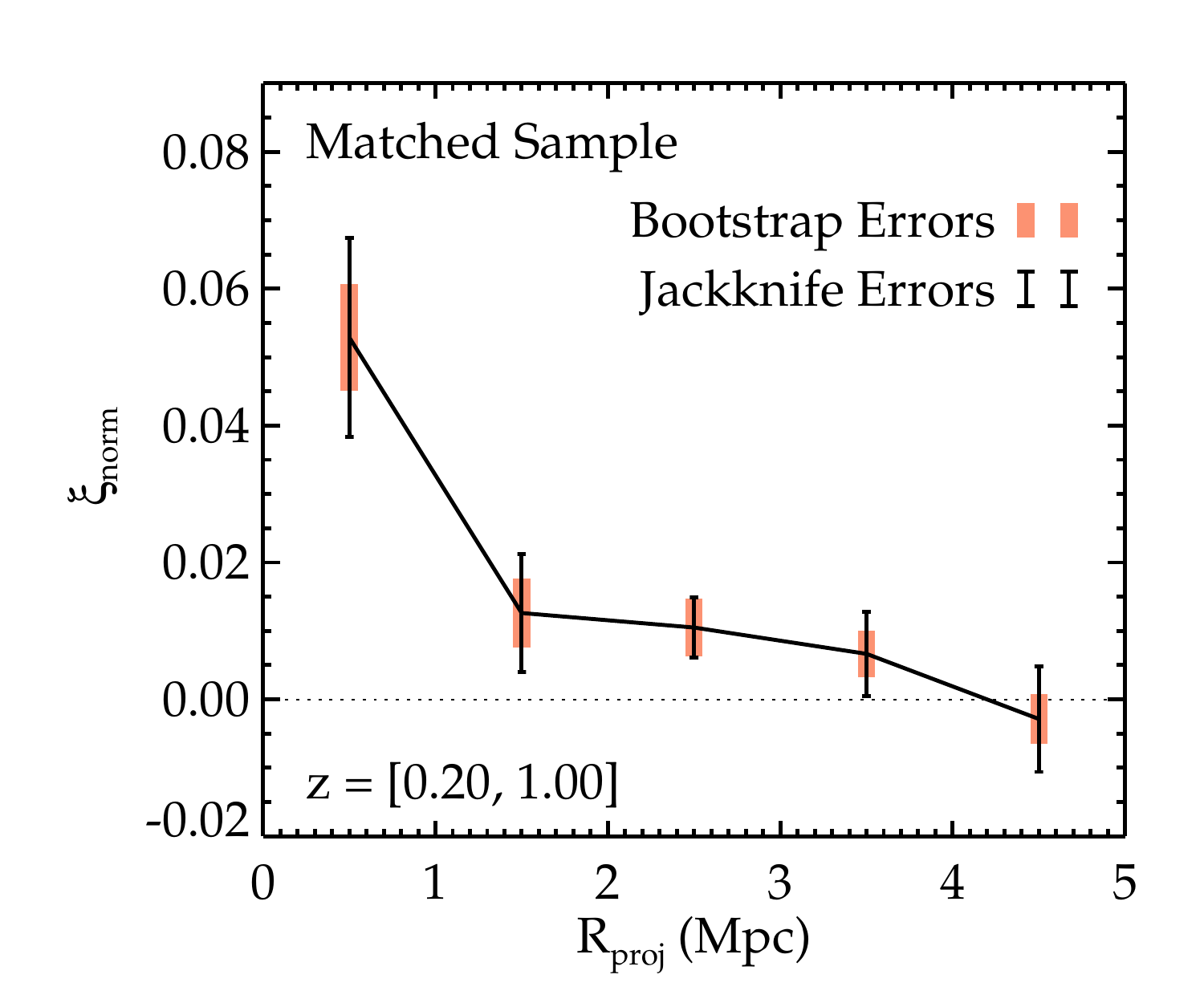}
  \caption{Normalized conformity signal, $\signorm$, for the matched sample to ${\Rproj<5}$~Mpc.  Both bootstrap (orange) and jackknife errors (black) are shown.  The jackknife errors exceed the bootstrap errors by up to a factor of $\sim2$.
}
  \label{fig:normsig_matched}
\end{figure}

Defining a normalized conformity signal serves two important purposes.  First, it enables us to clearly demonstrate the significant effects that cosmic variance can have on conformity measurements, which we show in Figures~\ref{fig:normsig_matched} and \ref{fig:normsig_fields_1halo} and discuss below in \S\ref{sec:errors} and \S\ref{sec:cosmic_var}.
Second, it is a metric we can use to \emph{quantitatively} compare the magnitude of the conformity signal we detect to the results of other conformity studies, as we do in \S\ref{sec:discussion} below.

Table~\ref{table:signal} presents the 
normalized conformity signal in the matched sample in integrated radial bins of 
$\Rproj=0$--1, {1--3}, and {3--5~Mpc}.
Over the full redshift range ${0.2<z<1.0}$ we find a normalized one-halo conformity 
signal of 5.3\% and a two-halo signal of 1.1\%.
We emphasize that galactic conformity is a very small effect, especially at two-halo scales, making it highly sensitive to observational uncertainty.
Galactic conformity therefore cannot be accurately measured without a sufficiently large sample volume.
The above measurements were made using over 60,000 galaxies in ${\sim2\times10^7}$ comoving Mpc$^3$ spanning over 5~Gyr of cosmic time.

In \S\ref{sec:SFR} above we note that the uncertainty of our stellar mass and SFR estimates introduces an uncertainty in the star-forming fraction of the full sample of $\sim1\%$.  To test how this error affects our measurement of $\signorm$ we recomputed the normalized conformity signal (Figure~\ref{fig:normsig_matched}) for the matched sample 10 times, each time drawing individual stellar masses and SFRs for each galaxy in the full sample (instead of using the median values) from a normal distribution with a width equal to the galaxy's stellar mass or SFR error.
The mean normalized one-halo conformity signal (0--1~Mpc) increased by $0.2\sigma$ and the two-halo signal (1--3~Mpc) increased by $1.4\sigma$.
This clearly indicates that the stellar mass and SFR errors are subdominant.

\subsection{Bootstrap Versus Jackknife Errors}\label{sec:errors}

In Table~\ref{table:signal} we estimate the uncertainty in $\signorm$ using both bootstrap and jackknife resampling, and quote the significance we find using each 
method as $\sigmaBS$ and $\sigmaJK$, respectively.
We compute bootstrap errors by selecting 90\% of the data randomly with replacement 200 times, and then taking the standard deviation of the 200 results.
To compute jackknife errors we divide the survey area of the matched sample into 10 regions of approximately $0.5~\degsq$ each.
We then compute $\signorm$ 10 times, systematically excluding one of the 10 jackknife samples each time, and take the standard deviation of the 10 results as the error.

Each method gives information about a different type of variation in our sample.
Bootstrap resampling provides an estimate of the variation of $\flate$ for the entire matched IP sample \emph{as a whole}.
It does not, however, take into account that our matched sample contains four spatially-distinct fields of different sizes on the sky.

Jackknife resampling estimates the uncertainty in $\flate$ \emph{due to field-to-field variation} (i.e.~cosmic variance) within the matched sample.
As seen in Table~\ref{table:signal}, jackknife resampling yields errors that are at least as large as bootstrap errors at all projected radii,
and which usually exceed bootstrap errors by a factor of $\sim2$.
Cosmic variance is therefore the dominate source of uncertainty in our result.

We emphasize that any meaningful measurement of conformity at $z>0.2$ should accurately account for cosmic variance by using multiple spatially-distinct fields and jackknife errors.
Bootstrap resampling is sufficient to estimate the uncertainty of a conformity signal \emph{within a single field}, but the result obtained with any one field cannot
realistically be extrapolated to draw conclusions about conformity on larger scales (see also \S\ref{sec:cosmic_var}).

Figure~\ref{fig:normsig_matched} shows $\signorm$ for the matched sample in ${d\Rproj=1}$~Mpc bins with both jackknife and bootstrap errors.
In the matched sample we find that for ${0<\Rproj<1}$~Mpc the bootstrap error of $\signorm$ is ${\pm0.008}$, which yields a significance of $\sigmaBS=6.8$, while the jackknife error is ${\pm0.015}$, with a significance of $\sigmaJK=3.6$.
 
The above result uses all star-forming and quiescent IPs in the matched sample, regardless of specific SFR (sSFR).  
To test whether the conformity signal is sensitive to the magnitude of the difference in sSFR between star-forming and quiescent IPs, we also measure one- and two-halo conformity with only the extreme high and low ends of the IP sSFR distribution.  Specifically, we compute $\signorm$ for the highest and lowest quartiles of IP sSFR (also matched in stellar mass and redshift distribution).

One-halo conformity over the full redshift range increases slightly to 5.5\%, while the uncertainty decreases to 1.2\%.
This increases $\sigmaJK$ to 4.7, even though the sample is half the size of the matched IP sample.
Two-halo conformity increases slightly to 1.5\%, but the uncertainty also increases to 0.9\%, which decreases $\sigmaJK$ to 1.7.

\subsection{Variance Between Fields}\label{sec:cosmic_var}

\begin{figure*}
  \epsscale{1.1}
  \epstrim{0.1in 0.3in 0.4in 0.3in}
  \plotone{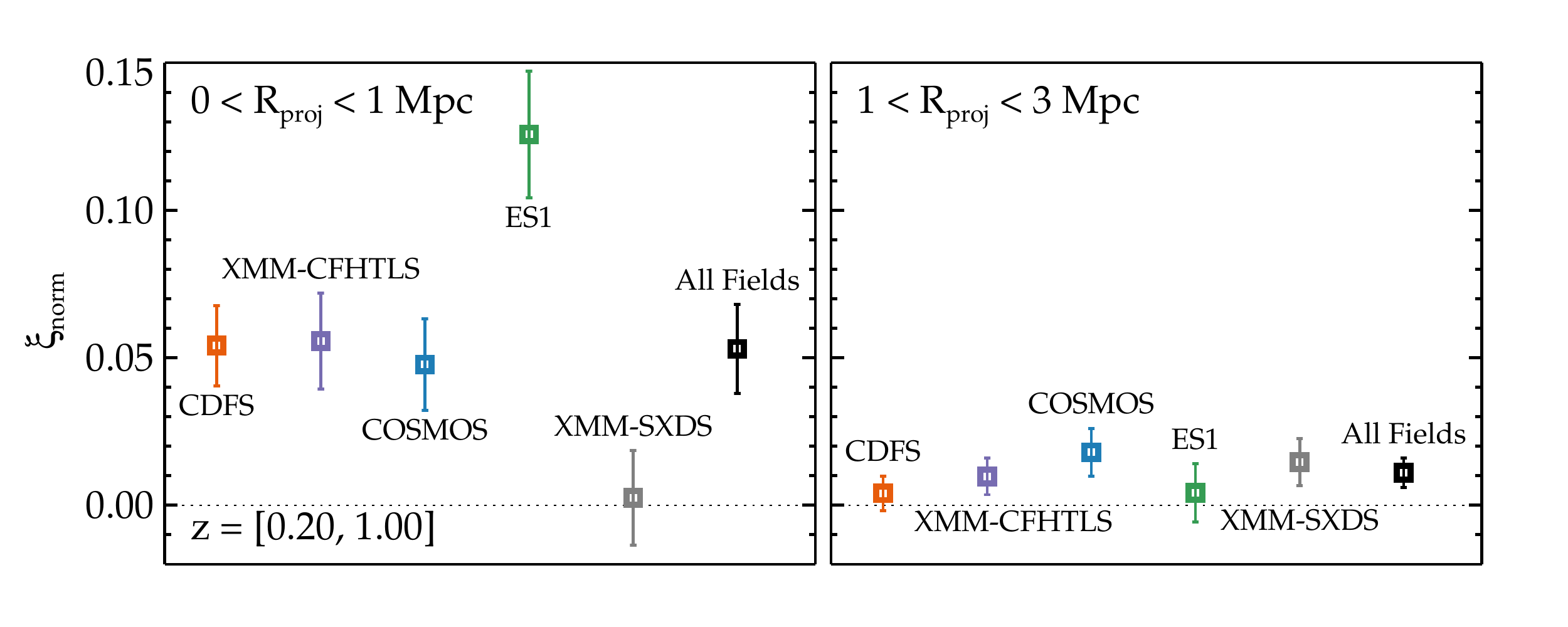}
  \caption{
One-halo ($\Rproj < 1$~Mpc, left) and two-halo ($1 < \Rproj < 3$~Mpc, right) conformity signals for each field and for the matched sample (``All Fields").  Individual field errors are estimated by bootstrap resampling within the field and the matched sample error is estimated by jackknife resampling.
}
  \label{fig:normsig_fields_1halo}
\end{figure*}

\input{tab03_signal_fields}

Errors estimated with jackknife resampling account for variation in the 
magnitude of the conformity signal among spatially-distinct regions of the sky.
The fact that $\sigmaJK$ is significantly less than $\sigmaBS$ for every conformity signal measurement in Table~\ref{table:signal}
illustrates the importance of accounting for cosmic variance in any conformity measurement.

We further investigate how the conformity signal in PRIMUS is sensitive to cosmic variance by measuring one- and two-halo conformity in each field individually.
The results are shown in Table~\ref{table:signal_fields} and Figure~\ref{fig:normsig_fields_1halo}.
The errors on the individual field measurements in Figure~\ref{fig:normsig_fields_1halo} are computed by bootstrap resampling within the field, and represent the uncertainty of the conformity signal \emph{in that field}.
The error on the signal measured over all five fields is computed by jackknife resampling from all fields, and represents the uncertainty of the conformity signal due to variation \emph{among} different fields.

The field-to-field variation within PRIMUS is substantial.
Among the five fields in the matched sample the one-halo conformity signal varies from over $12\%$ with $\sigmaBS=5.9$ in ES1, to $\sim5$\% in 
CDFS, COSMOS, and XMM-CFHTLS, to $0\%$ with $\sigmaBS\simeq0$ in XMM-SXDS.
The two-halo signal has similar field-fo-field variation relative to its smaller overall magnitude of $\sim1\%$.
This variation clearly indicates the importance of measuring conformity in multiple fields.  A large dispersion exists in the strength of conformity among PRIMUS fields, and the signal in any one field can differ significantly from the mean.

\subsection{Redshift and Stellar Mass Dependence}\label{sec:z_mass_bins}

\begin{figure*}
  \epsscale{1.0}
  \epstrim{0.2in 0.3in 0.4in 0.8in}
  \plotone{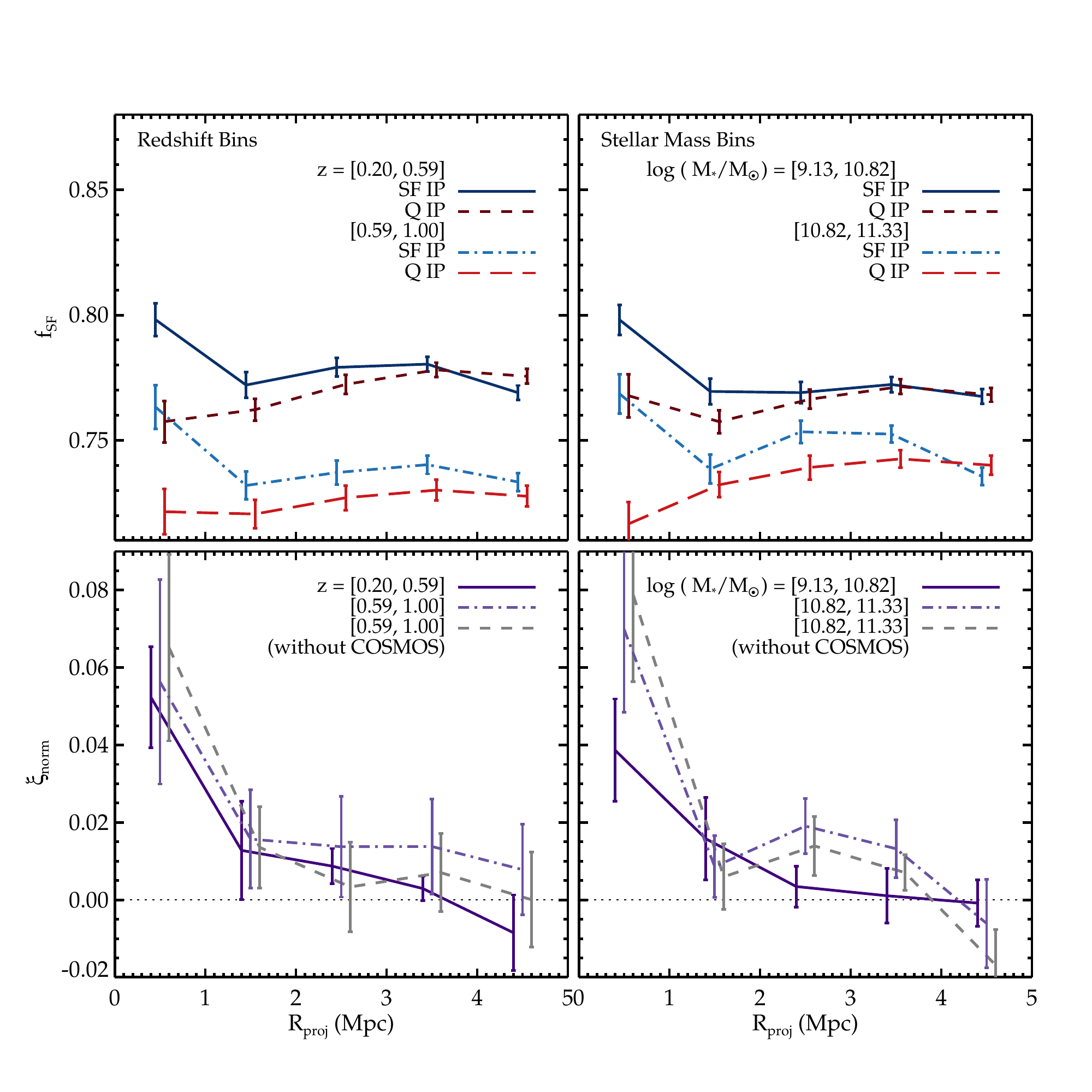}
  \caption{
Top panels: Star-forming neighbor fractions for star-forming (solid and dash-dot blue lines) and quiescent (dashed red lines) IPs in our matched sample divided into two redshift bins (left) and two stellar mass bins (right).  Errors are computed by bootstrap resampling and are offset for clarity.
Bottom panels: $\signorm$ for the corresponding redshift and stellar mass divisions in the top panels.  Errors are computed by jackknife resampling.
The bottom panels also show $\signorm$ for the higher redshift bin (left) and higher stellar mass bin (right) computed \emph{without} the COSMOS field (dashed gray line).
}
  \label{fig:latefrac_normsig_compare}
\end{figure*}

He16 predicts that conformity strength (specifically two-halo) should decrease with both increasing central galaxy halo mass and increasing redshift, weakening significantly by $z\sim1$ and disappearing entirely by $z\sim2$.
With PRIMUS we can test for trends in conformity signal strength with redshift to $z=1$, and with halo mass using IP stellar mass as a proxy for halo mass.
We further divide the matched sample into two redshift bins and two stellar mass bins 
to investigate the dependence in the magnitude of the signal on redshift or stellar mass.
In Figure~\ref{fig:latefrac_normsig_compare} we divide the matched IP sample into two redshift bins, ${0.2<z<0.59}$ and ${0.59<z<1}$, and two stellar mass bins, 
${9.13<\logM<10.82}$ and ${10.82<\logM<11.33}$, each containing equal numbers of IPs.
The upper panels show $\flate$ for star-forming and quiescent IPs in each redshift or stellar mass bin, while the lower panels plot the corresponding values of
$\signorm$ for each radial bin. The normalized signal and significance are given in 
Table~\ref{table:signal}.

When dividing into redshift bins the one-halo conformity signal in both bins is comparable to the 5.3\% signal observed over the full redshift range.
The significance of the ``low'' redshift (${0.2<z<0.59}$) one-halo conformity signal increases to 
${\sigmaJK=4.0}$, while the significance of the ``high'' redshift one-halo conformity signal drops to ${\sigmaJK=2.1}$.
The magnitude of the two-halo conformity signal in each bin also remains comparable to the full redshift range signal of 1.1\%, but the uncertainty in each bin also increases,
reducing $\sigmaJK$ from 2.5 for the full redshift range to 1.7 and 1.6 for the low and high redshift bins, respectively.

Dividing the matched sample into two stellar mass bins containing equal numbers of IPs, we find a significant one-halo conformity signal in both stellar mass bins:~$\sim4$\% in the lower-mass bin ($9.13<\logM<10.82$), and $\sim7$\% in the higher-mass bin ($10.82<\logM<11.33$).
We also find that two-halo conformity is stronger at higher than at lower stellar mass, although the significance of this trend is low due to the large error of our lower-mass two-halo signal.

While ideally one would use narrower stellar mass bins to test the stellar mass dependence of the conformity signal, our sample size restricts us to a lower-mass binwidth of 1.7~dex.
If conformity is intrinsically stronger at lower stellar mass, averaging over too wide a range in stellar mass would dilute a strong signal contribution from lower masses.

Assuming that galaxy stellar mass is tightly coupled with host halo mass, our results appear to contradict the He16 prediction that conformity strength should decrease with increasing halo mass.
However, the He16 prediction is specifically for two-halo conformity, where we do not have significant results as a function of stellar mass.
Therefore, we can not test the two-halo stellar mass dependence of conformity with our sample.

\subsubsection{The Effect of the COSMOS Field}\label{sec:cosmos}

The COSMOS field contains substantial large-scale structures at $z\sim0.35$ and $z\sim0.7$ \citep[e.g.,][]{McCracken07, Meneux09, Kovac10}, presenting another opportunity to test the impact of field-to-field variation on our results.
As Figure~\ref{fig:normsig_fields_1halo} and Table~\ref{table:signal_fields} show, one-halo conformity in the COSMOS field alone agrees well with the one-halo signal in the matched sample.
However, we also measure two-halo conformity for each field individually, and find that it is stronger in COSMOS than in any other field.
Additionally, when we divide the matched sample into two redshift and two stellar mass bins, two-halo conformity is stronger in both the higher redshift and higher stellar mass bin (see Figure~\ref{fig:latefrac_normsig_compare} and Table~\ref{table:signal}).

To investigate the degree to which COSMOS contributes to the higher redshift and higher stellar mass two-halo conformity signal we recomputed $\signorm$ for these bins using all of the matched sample \emph{except} COSMOS.
The result is shown in the lower panels of Figure~\ref{fig:latefrac_normsig_compare} (dashed gray lines).
In both cases (higher redshift and higher stellar mass) two-halo conformity \emph{without} COSMOS is weaker than the result for all fields.
However, the results including and excluding COSMOS are each within the uncertainty of the other for both the higher stellar mass and higher redshift bins.

We conclude that the $1.5~(\pm0.5)\%$ two-halo conformity signal we observe at higher stellar mass (${10.8\lesssim \logM
 \lesssim11.3}$) is not dominated by a single field, but it is likely inflated by COSMOS.
The two-halo signal strength trend with stellar mass we observe is \emph{less} at odds with He16's predictions if the COSMOS field is excluded.
However, this is not a statement about COSMOS specifically, but about the degree to which conformity measurements are sensitive to cosmic variance in general.
Surveys larger than the 5.5~$\degsq$ of our matched sample, with comparable depth and sampling density, are required
to confidently test existing predictions about the relat onship between conformity strength and both redshift and mass.

\subsection{The Relationship Between IP Quenching and Environment}\label{sec:environment}

If two-halo conformity is due to large-scale tidal fields and adjacent halos being similarly affected by larger-scale overdensities, then we would expect to observe a correlation between the quenched fraction of central galaxies and large-scale environment.

Behroozi et al.~(in prep., hereafter~\citePB) measure central galaxy quenched fraction versus environment for a sample of SDSS galaxies at ${0.01<z<0.057}$ with ${10<\logM<10.5}$.
Within this sample, \citePB define ``central'' galaxies as those with no larger (in stellar mass) neighbors within a projected distance of 500~kpc and 1000~\kms in redshift.
\citePB define neighbors as galaxies of stellar mass $\mneigh$ where ${0<(\mcentral-\mneigh)<0.5}$~dex, and within a projected distance of 0.3--4~Mpc and 1000~\kms in redshift from the central galaxy.
These cuts reduce any correlation between environment and central galaxy stellar mass, in that the median stellar mass of the central galaxies is only very weakly, if at all, correlated with environment.
\citePB find that the star-forming fraction of central galaxies is negatively correlated with environment, decreasing by about a factor of two as the number of neighbors increases by an order of magnitude from $\sim10$ to $\sim100$.
\citePB also find that the mean sSFR of \emph{star-forming} central galaxies does \emph{not} depend on environment.

We test for the same relationships in PRIMUS by comparing the star-forming fraction with environment for a subset of isolated primaries.
To ensure that both our IP and neighbor samples are complete, we consider only IPs with stellar masses 0.5~dex \emph{greater} than the completeness limits described in \S\ref{sec:mass_limit}.
We use the same definition of neighbors as \citePB, except we use ${\Delta z = 2\,\sigmaz}$ instead of \citePB's 1000~\kms to account for our sample's larger uncertainty in redshift.
At the redshift range of our sample ${2\,\sigmaz \approx 3000}$~\kms.
For accurate environment measurements our neighbor sample must be complete to 0.5~dex below the minimum IP stellar mass for a particular redshift range, field, and galaxy type.
Following \citePB, our measure of environment is \Nneigh, the sum of the statistical weights (see \S\ref{sec:targ_weight}) of all neighbors of an IP galaxy, which need not be an integer.

We select IPs in three bins in stellar mass, each of which spans ${0.2<z<\zmax}$:~${\log\,(\mIP/\msun)=10.1}$~to~10.4 (${\zmax=0.65}$), 10.4~to~10.7 (${\zmax=0.8}$), and 10.7~to~11.0 (${\zmax=1.0}$).
These bins are narrower than the 0.5~dex width used by \citePB because the PRIMUS mass completeness limits depend strongly on redshift; for $z>0.65$ {($z>0.8$)} our neighbor sample is only complete for IP masses greater than $10^{10.4}$ ($10^{10.7}$)~$\msun$.
Narrow bins allow us to measure the relationship between IP quenched fraction and environment over the full PRIMUS redshift range.

\begin{figure*}
  \epsscale{1.0}
  \epstrim{0.6in 0.3in 0.7in 0.8in}
  \plotone{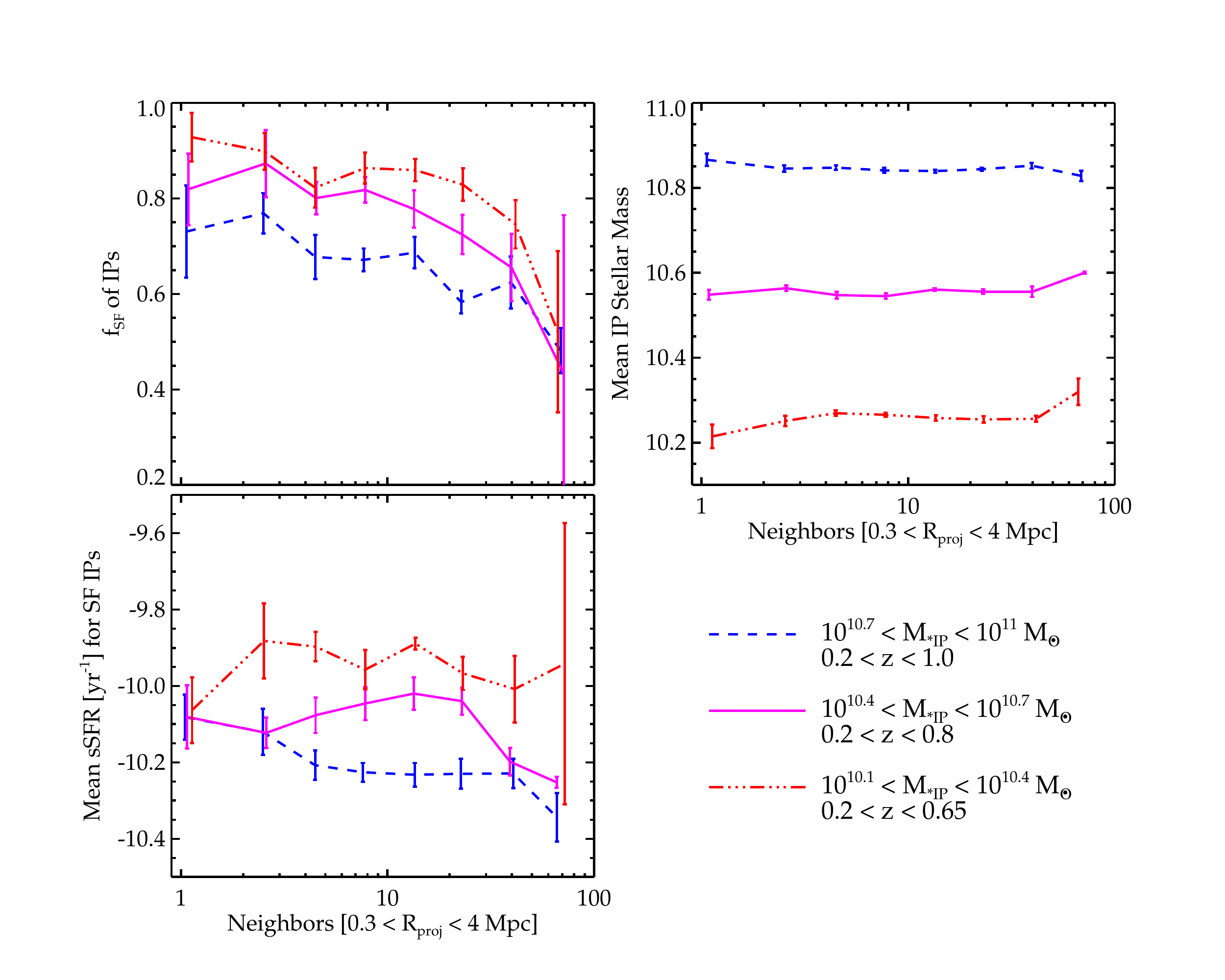}
  \caption{
Star-forming fraction of IPs (top left), median IP stellar mass (top right), and mean sSFR of star-forming IPs (bottom left),
each as a function of environment for three bins in IP stellar mass:~${10.1<\log\,(\mIP/\msun)<10.4}$ (dash-dot red line),
${10.4<\log\,(\mIP/\msun)<10.7}$ (solid magenta line), and
${10.7<\log\,(\mIP/\msun)<11.0}$ (dashed blue line).
Neighbors are defined as galaxies of stellar mass $\mneigh$, where ${0<(\mIP-\mneigh)<0.5}$~dex, within ${0.3<\Rproj<4}$~Mpc and $2\,\sigmaz$ in redshift space from the IP.
Errors are computed by jackknife resampling.
}
  \label{fig:environment}
\end{figure*}

Figure~\ref{fig:environment} shows the star-forming fraction of IPs (top left), median IP stellar mass (top right), and mean sSFR for star-forming IPs (bottom left), each as a function of environment for three bins in IP stellar mass.
The top right panel is a check that the IP stellar mass distribution within each bin is independent of environment:~as shown, IPs with few as well as many neighbors have the same stellar mass.
Similarly, the bottom left panel shows only weak correlation between the sSFR of star-forming IPs and environment, and no correlation for the lowest mass bin.  This clearly indicates that as long as a galaxy is forming stars, the sSFR is not strongly (if at all) dependent on the large-scale environment of the galaxy.

However, for all three stellar mass bins, the star-forming fraction of IPs is roughly constant for ${\Nneigh\lesssim10}$, then falls off as $\Nneigh$ increases.
The difference in IP star-forming fraction between IPs with ${\Nneigh<10}$ and ${\Nneigh>30}$ is
$\sim13$\% ($2.1\sigma$) for $10.1<\log\,(\mIP/\msun)<10.4$,
$\sim20$\% ($3.4\sigma$) for $10.4<\log\,(\mIP/\msun)<10.7$, and
$\sim10$\% ($1.8\sigma$) for $10.7<\log\,(\mIP/\msun)<11.0$.
The difference is statistically significant only for the middle stellar mass bin.

We can increase the signal-to-noise in this measurement by considering the decrease in IP star-forming fraction between IPs with
${\Nneigh<10}$ and ${\Nneigh>30}$ for wider stellar mass bins. 
For ${10.1<\log\,(\mIP/\msun)<10.7}$ (${\zmax=0.65}$) the IP star-forming fraction decreases by $\sim15$\% ($4.9\sigma$), 
while for ${10.4<\log\,(\mIP/\msun)<11.0}$ (${\zmax=0.8}$) we find a decrease of $\sim11$\% ($2.9\sigma$).

Two conclusions can be drawn from Figure~\ref{fig:environment}.
The first is that \emph{at fixed stellar mass} central galaxies are more likely to be quenched in large-scale ($\sim4$~Mpc) environments.
This may be a consequence of the known correlations between 
halo mass and greater quenched fraction, and between halo mass and 
large-scale environment.

The second conclusion is not due to known correlations, and it is that as long as a central galaxy in a dense environment is forming stars, it does so \emph{as efficiently} as a star-forming central galaxy of the same stellar mass in a low-density environment.

These results are consistent with \citePB and indicate that the higher probability that a central galaxy is quenched when residing in a large-scale overdensity persists to $z\sim0.5$--1.  This measurement can also be made at higher significance than the usual ``conformity'' signal (as presented above). 

\begin{figure}
  \epsscale{1.1}
  \epstrim{0in 0.1in 0.4in 0.7in}
  \plotone{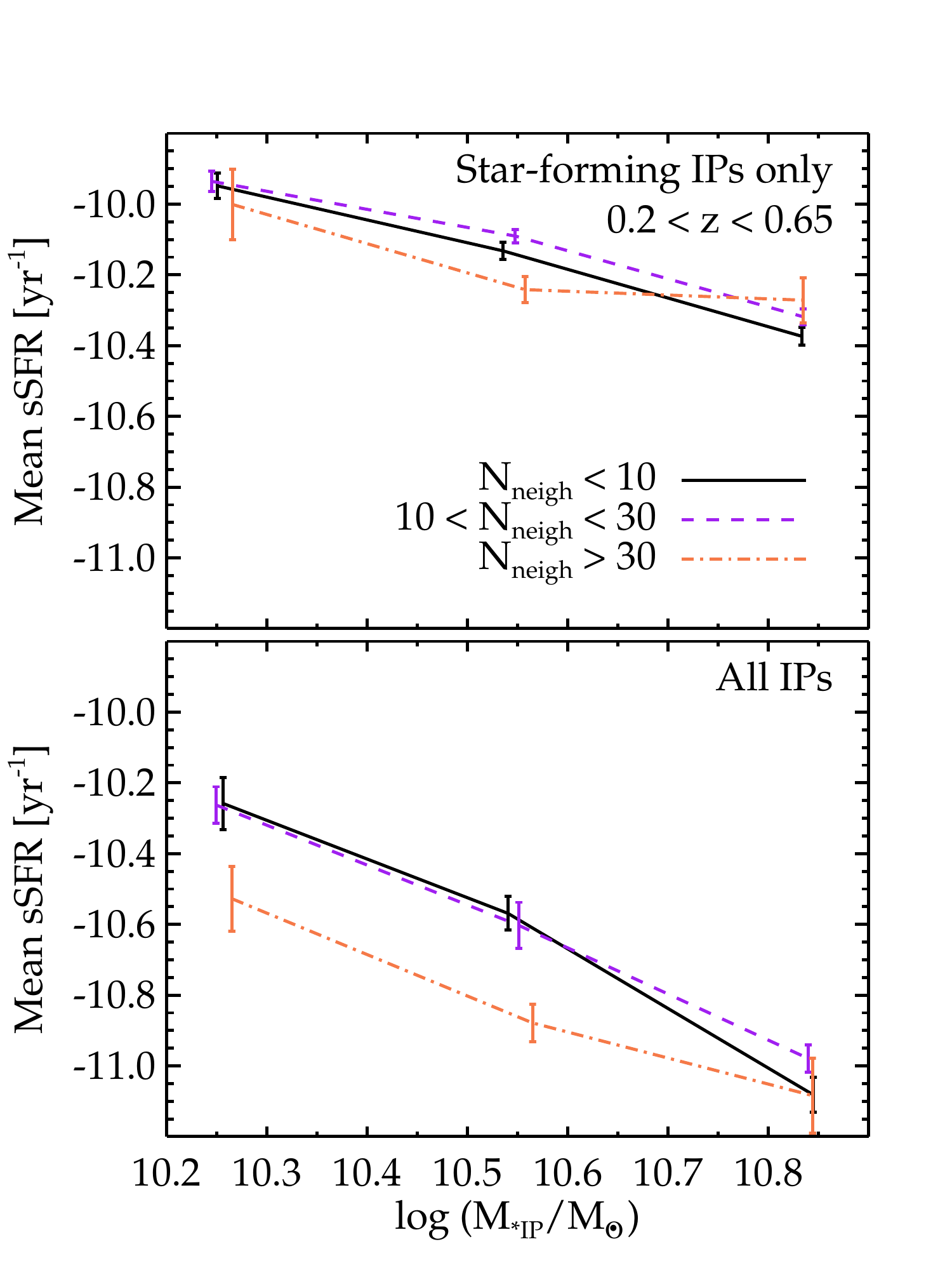}
  \caption{
Mean sSFR for star-forming (top panel) and all (bottom panel) IP galaxies as a function of stellar mass for three bins in environment:~${\Nneigh<10}$ (solid black line), ${10<\Nneigh<30}$ (dashed purple line), and ${\Nneigh>30}$ (dash-dot orange line). Errors are computed by jackknife resampling.
}
  \label{fig:sSFR_vs_mstar}
\end{figure}

To further investigate the relationship between central galaxy sSFR and environment, in Figure~\ref{fig:sSFR_vs_mstar} we plot the
mean sSFR for all IPs and for star-forming IPs only as a function of stellar mass in three bins of environment,
${\Nneigh<10}$, ${10<\Nneigh<30}$, and ${\Nneigh>30}$, in the mass range ${10^{10.1}<\mIP<10^{11.0}~\msun}$ and redshift range ${0.2<z<0.65}$.

Specific SFR is negatively correlated with IP stellar mass in all three environment bins.
This trend is highly significant (${\ge4\sigma}$) both for all IPs and for star-forming IPs alone,
with the exception of the ${\Nneigh>30}$ bin of star-forming IPs, where ${\sigma\sim2.3}$.
This bin also contains the fewest galaxies, which likely contributes to the lower significance.

There is no statistical difference between the low and intermediate density bins, although this could be a result of our inability to robustly measure environment.
Additionally, there are no statistical differences among the three environment bins for IPs of high stellar mass ($\mIP\gtrsim10^{10.7}~\msun$),
again for both all IPs and star-forming IPs only.

Considering just the bottom panel of Figure~\ref{fig:sSFR_vs_mstar}, a statistically significant difference of $\sim0.3$~dex in sSFR
does exist between low- and intermediate-mass IPs (${10^{10.1}<\mIP<10^{10.7}~\msun}$) in very dense environments (${\Nneigh>30}$) and those with ${\Nneigh<30}$.
At higher stellar mass ($\mIP\gtrsim10^{10.7}~\msun$) and within the errors, 
\emph{as long as a central galaxy is forming stars} its rate of star formation is \emph{not} influenced by large-scale environment.
However, overdense environments are clearly correlated with central galaxy quenching.

%% file: tab04_signal_matched.tex
\setlength{\tabcolsep}{0.05in}
\begin{deluxetable*}{ccrrrcccrcccrcc}
\tablecaption{Normalized conformity signal $\signorm$ for the matched sample, two redshift bins, and two stellar mass bins for three ranges of projected radius: $\Rproj=0$--1, 1--3, and 3--5~Mpc.  Errors shown are computed by jackknife resampling.  The significance of all signals is shown for both jackknife ($\sigmaJK$) and bootstrap ($\sigmaBS$) errors.  \label{table:signal}}
\tablewidth{0pt}
\tablehead{
\multicolumn{2}{c}{$N_{\textrm{IP}}$} & \colhead{} & \colhead{} &
\multicolumn{3}{c}{$0.0 < R < 1.0$~Mpc} & {} &
\multicolumn{3}{c}{$1.0 < R < 3.0$~Mpc} & {} &
\multicolumn{3}{c}{$3.0 < R < 5.0$~Mpc} \\
\cline{1-2}\cline{5-7}\cline{9-11}\cline{13-15} \\
\colhead{SF\tablenotemark{a}} & \colhead{Q} &
\multicolumn{1}{c}{$z$\tablenotemark{b}} &
\multicolumn{1}{c}{$\log\,(\mstar/\msun)$\tablenotemark{c}} &
\colhead{$\signorm$} & \colhead{$\sigmaJK$} & \colhead{($\sigmaBS$)} & {} &
\colhead{$\signorm$} & \colhead{$\sigmaJK$} & \colhead{($\sigmaBS$)} & {} &
\colhead{$\signorm$} & \colhead{$\sigmaJK$} & \colhead{($\sigmaBS$)} \\
\cline{1-15} \\
\multicolumn{15}{c}{Matched Sample}
}
\startdata
$4,185$ &
$6,197$ &
[0.20, 1.00] &
[9.13, 11.33] &
$0.053\pm0.015$ & 3.6 & $(6.8)$  & {} &
$0.011\pm0.005$ & 2.5 & $(3.5)$  & {} &
$0.002\pm0.005$ & 0.3 & $(0.6)$  \\
\cutinhead{Redshift Bins}
$2,241$ &
$3,096$ &
[0.20, 0.59] &
[9.13, 11.25] &
$0.052\pm0.013$ & 4.0 & $(4.9)$  & {} &
$0.010\pm0.006$ & 1.7 & $(2.5)$  & {} &
$-0.003\pm0.005$ & 0.7 & $(1.1)$  \\
$1,945$ &
$3,101$ &
[0.59, 1.00] &
[10.11, 11.33] &
$0.056\pm0.026$ & 2.1 & $(4.5)$  & {} &
$0.015\pm0.009$ & 1.6 & $(2.8)$  & {} &
$0.011\pm0.008$ & 1.2 & $(2.7)$  \\
\cutinhead{Stellar Mass Bins}
$2,385$ &
$3,069$ &
[0.20, 1.00] &
[9.13, 10.82] &
$0.039\pm0.013$ & 2.9 & $(3.7)$  & {} &
$0.008\pm0.006$ & 1.5 & $(1.9)$  & {} &
$0.000\pm0.005$ & 0.0 & $(0.0)$  \\
$1,801$ &
$3,128$ &
[0.20, 1.00] &
[10.82, 11.33] &
$0.070\pm0.021$ & 3.3 & $(5.9)$  & {} &
$0.015\pm0.005$ & 2.8 & $(3.0)$  & {} &
$0.003\pm0.007$ & 0.4 & $(0.8)$  \\
\enddata
\tablenotetext{a}{Number of \emph{unique} SF IPs. SF IPs are upweighted such that the sum of the weights equals the total number of Q IPs.}
\tablenotetext{b}{Redshift range of IP (sub)sample.}
\tablenotetext{c}{Stellar mass range of IP (sub)sample}
\end{deluxetable*}

%% file: tab03_signal_fields.tex
\setlength{\tabcolsep}{0.1in}
\begin{deluxetable}{lrrrr}
\tablecaption{Significance of the one-halo conformity signal in individual fields.
\label{table:signal_fields}
}
\tablewidth{0pt}
\tablehead{
\colhead{Field} & \colhead{$N_{\textrm{IP}}$\tablenotemark{a}} & \colhead{$N_{\textrm{SF-IP}}$\tablenotemark{b}} & \colhead{$N_{\textrm{Q-IP}}$} & \colhead{$\sigmaBS$\tablenotemark{c}}
}
\startdata
CDFS &
$2,837$ &
$1,139$ &
$1,698$ &
$4.0$ \\
COSMOS &
$1,830$ &
$731$ &
$1,099$ &
$3.1$ \\
ES1 &
$1,011$ &
$390$ &
$621$ &
$5.9$ \\
XMM-CFHTLS &
$3,222$ &
$1,325$ &
$1,897$ &
$3.4$ \\
XMM-SXDS &
$1,482$ &
$600$ &
$882$ &
$0.2$ \\
\cline{1-5} \\
Matched Sample\tablenotemark{c} &
$10,392$ &
$4,185$ &
$6,207$ &
$3.6$ \\
\enddata
\tablenotetext{a}{Total number of unique IPs.}
\tablenotetext{b}{Number of unique SF IPs.}
\tablenotetext{c}{The significance of the matched sample is computed using jackknife errors.}
\end{deluxetable}

%% file: discussion.tex
\section{Discussion}\label{sec:discussion}

We have presented a significant detection of both one-halo and two-halo galactic conformity 
at ${0.2 < z < 1.0}$ using the largest faint galaxy spectroscopic redshift survey completed to date.  
Ours is currently the only study of galactic conformity at intermediate redshift performed with 
spectroscopic redshifts, and is the first detection of two-halo conformity at $z>0.2$.
In this section we compare our results with existing conformity studies,
both at low ($z<0.2$) and higher (${0.2<z<2.5}$) redshift,
and discuss the physical implications of our results.

\subsection{Comparison to Previous Low Redshift Studies}\label{sec:compare_low}

The original discovery of conformity, \citet{Weinmann06}, measured the star-forming satellite fraction for quiescent and
star-forming central galaxies at fixed halo mass.
Their estimates of the star-forming satellite fraction range from $\sim20\%$ to $\sim65\%$ for quiescent centrals, and $\sim45\%$ to $\sim80\%$ for star-forming
centrals, depending on halo mass, central galaxy luminosity, and whether galaxy type was determined by color or by sSFR.
If we apply Equation~\ref{eq:signorm} to these star-forming fractions, the magnitude of the one-halo signal found by \citet{Weinmann06} is $\sim20\%$ at $\mhalo \sim10^{12}\msun$ (roughly corresponding to $\mstar \sim10^{10}\msun$) compared to the 5.3\% 
we find at higher redshift, which is qualitatively consistent with the He16 prediction that conformity strength decrease with increasing redshift.

We note that if we define the normalized conformity signal (Equation~\ref{eq:signorm}) to be in terms of quiescent neighbor fraction instead of star-forming fraction then the \citet{Weinmann06} signal is $\sim35\%$ at $\mhalo \sim10^{12}\msun$ while our signal would be $\sim17\%$.  Because the overall star-forming fraction of galaxies in our sample is $\sim75\%$ a difference in star-forming neighbor fraction between star-forming and quiescent IPs of 5\% is smaller \emph{relative to the mean star-forming fraction} than to the mean quiescent fraction.  However the fundamental result is unchanged:~\citet{Weinmann06} find a stronger conformity signal at $z<0.2$ than the signal we find at higher redshift.  We emphasize that throughout this work the physical interpretation of our results does \emph{not} depend on whether the normalized conformity signal is defined in terms of star-forming or quiescent neighbor fraction.

We also compare our results with those of K13, whose methodology for defining isolated central galaxies is used in this work.
Unlike \citet{Weinmann06} and our study, K13 did not measure the star-forming fraction, but instead 
compared the median satellite galaxy sSFR for quartiles of isolated primary (i.e. ``central'') galaxy sSFR at fixed stellar mass.
K13 found a significant galactic conformity signal across the full central 
galaxy stellar mass range studied (${5\times10^9~\msun}$ to ${3\times10^{11}~\msun}$) in a sample of 
SDSS galaxies with ${0.017 < z < 0.03}$.
Our main result of $\gtrsim3\sigma$ detections of one- and two-halo conformity at ${0.2<z<1}$ is consistent with the signal K13 find at lower redshift.

K13 also compared low-mass ($9.7 < \logM < 10.3$) and high-mass ($10.7 < \logM < 11.5$) samples of 
central galaxies and found that the scale dependence of the conformity signal depends on the central 
galaxy stellar mass.  Specifically, K13 find that at low redshift two-halo conformity exists for 
low-mass central galaxies, while for high-mass central galaxies conformity is confined to one-halo scales.

Contrary to K13, within PRIMUS we do not detect significant differences in the measured conformity 
signal with stellar mass;
however, as discussed above, the error bars on our measurements may be too large to detect such a signal.
Additionally, the stellar mass ranges we study differ from those of K13.
To keep our sample sizes large and minimize uncertainty, our low-mass bin spans 1.7~dex from 
${9.1 \lesssim \logM \lesssim 10.8}$, which is a much wider range than in K13.
Our high-mass bin (${10.8 \lesssim \logM \lesssim 11.3}$) spans only 0.5~dex, and is a subset of the high-mass bin in K13.

As mentioned above, K13 compare quartiles of central galaxy sSFR instead of using a binary classification of galaxies as either star-forming or quiescent, as we do here.
However, as discussed in \S\ref{sec:errors} above, we do not find different results within our sample if we compare quartiles in sSFR instead of using a binary galaxy type classification.
Indeed, as we showed in \S\ref{sec:environment}, what appears to be driving the conformity signal is whether a galaxy is indeed quenched.
Therefore, using quartiles in sSFR or a binary classification should yield similar results.

K13 find a two-halo conformity signal to a projected distance of $\sim4$~Mpc, while the two-halo signal we measure disappears by $\sim3$~Mpc.
This is consistent with the prediction of He16 that the scale dependence of conformity (i.e., the relative signal strength at a given distance) should weaken with increasing redshift, and also supports the idea that galactic conformity is an indirect result of large-scale tidal fields.
However, we note that in PRIMUS we have larger error bars than in SDSS, which could make it more difficult to detect a signal on larger scales.

\subsection{Comparison to Previous Higher Redshift Studies}\label{sec:compare_high}

We now compare our results with the two existing studies of conformity at higher redshift, H15 and \citet{Kawinwanichakij16}.

H15 used photometric redshifts to search for one-halo conformity at ${0.4<z<1.9}$ in the ${0.77~\degsq}$ 
UKIRT Infrared Deep Sky Survey \citep[UKIDSS;][]{Lawrence07} Ultra Deep Survey (UDS) field, which overlaps with our XMM-SXDS field.
They estimated the redshift uncertainty of their sample to be ${0.014 \lesssim \sigmaz \lesssim 0.088}$ and corrected for background contamination using the method described in \citet{Chen06}.

H15 defined central galaxies as those with no other galaxies within 450~projected kpc and ${\sqrt{2}\sigmaz(1 + z)}$
that have stellar mass more than 0.3~dex (their expected uncertainty in stellar mass) greater than the mass of the central galaxy.
Instead of star-forming (or quiescent/passive) \emph{fractions} of satellite galaxies,
H15 measured the radial density profiles (number per kpc$^2$) of quiescent and all satellite galaxies for mass-matched samples of quiescent and star-forming central galaxies in logarithmic radial bins to a projected distance of 1~Mpc, and claim to detect one-halo conformity at $>3\sigma$ to $z\sim2$.

By our definition the normalized conformity signal (Equation~\ref{eq:signorm}) H15 find is an order of magnitude larger than our one-halo result.
(As in \S\ref{sec:compare_low} we note that if we define the normalized conformity signal to be in terms of quiescent neighbor fraction instead of star-forming fraction then the H15 signal exceeds 100\% while our signal is $\sim17\%$. The fundamental result that the magnitude of H15's one-halo signal is much larger than ours is unchanged.)
This discrepancy is especially puzzling considering that H15's field, XMM-SXDS, is the only field in which we measure \emph{no} one-halo conformity, although our redshift range only partially overlaps with theirs.
Given the large uncertainties of their photometric redshifts, interlopers likely have a significant effect on H15's results.
For example, as H15 note, because they count and background-correct their quiescent and all satellite galaxy samples separately they in some cases obtain quiescent satellite fractions that are negative or greater than unity.

Using photometric redshifts from three surveys totaling 2.37~$\degsq$,
UltraVISTA \citep{McCracken12},
UKIDSS \citep{Lawrence07} UDS (Almaini et al., in prep.),
and the FourStar Galaxy Evolution Survey \citep[ZFOURGE;][]{Spitler12},
\citet{Kawinwanichakij16} tested for one-halo conformity in four redshift bins over the range ${0.3 < z < 2.5}$ for central galaxies with ${\mstar>10^{10.5}~\msun}$.
They defined central galaxies as those without any more massive galaxies within a projected distance of 300 comoving kpc (ckpc).
Satellite galaxies were defined as those with ${\mstar>10^{10.2}~\msun}$ and a redshift difference of ${\Delta z \le 0.2}$ from a the central galaxy.
\citet{Kawinwanichakij16} estimated the average quiescent fraction of satellite galaxies within 300 projected ckpc for stellar mass-matched samples of quiescent and star-forming central galaxies.
They did \emph{not} match the redshift distributions of their quiescent and star-forming central galaxy samples because the difference between the mean redshifts of these two samples is comparable to the redshift uncertainty (${0.01 \lesssim \sigmaz \lesssim 0.05}$) in each redshift interval they studied.

We can compare our results with those of \citet{Kawinwanichakij16} at ${0.3<z<0.6}$ and ${0.6<z<0.9}$, where they claim 
``less significant'' ($1.4\sigma$) and ``strong'' ($4.5\sigma$) detections, respectively.
Our results broadly agree over both redshift intervals combined, but in terms of significance we find the opposite:~our one-halo conformity signal has ${\sigmaJK=4}$ at ${0.2<z<0.59}$, and only ${\sigmaJK=2.1}$ at ${0.59<z<1}$.

The \emph{magnitude} of the observed conformity effect is quite different when measured with photometric versus spectroscopic redshifts.
H15 found a difference in raw quiescent fractions for quiescent and star-forming central galaxies of up to $\sim50$--60\% in their lower redshift bin ($0.4<z<1.3$).
In \citet{Kawinwanichakij16} the difference is as much as $\sim10\%$ at ${0.3<z<0.6}$ and up to $\sim30\%$ at ${0.3<z<0.6}$.
These numbers correspond to normalized conformity signals (defined by Equation~\ref{eq:signorm}) that are \emph{at least an order of magnitude} greater than the ${\sim5\%}$ one-halo conformity signal we find with spectroscopic redshifts alone.
Even more puzzling is that the larger uncertainties of photometric redshifts would be expected to \emph{dilute} a conformity signal, not enhance the effect.

Other factors could affect the measured star-forming satellite fractions of star-forming and quiescent primary galaxies, including differences in both central and satellite galaxy selection criteria, as well as how galaxies are classified as either star-forming or quiescent.
In the former case, both \citet{Kawinwanichakij16} and especially H15 use isolation criteria different from ours to select primary galaxies and their satellites; they adopt smaller projected radii, larger $\sigmaz$, and less conservative stellar mass limits on galaxies within the spatial boundary for isolation.

H15 and \citet{Kawinwanichakij16} both used a cut in rest-frame ${V-J}$ versus ${U-V}$ color to divide their samples into star-forming and quiescent galaxies, while Br16b used a redshift-dependent cut in $M_g$ versus ${(u-g)}$ color to divide their photometric sample.
As galaxy type distributions are bimodal for a variety of parameters, the precise method used to divide a sample should not have a strong effect on the outcome, provided the estimates of the parameters used (color, sSFR, etc.) are robust.

Br16b performed a complimentary study to ours using 
cross-correlation measurements between the PRIMUS spectroscopic and (deeper) photometric galaxy samples.
Specifically, they measured the overdensities of quiescent PRIMUS photometric galaxies within a physical deprojected distance of $\sim1/h$~Mpc using PRIMUS spectroscopic galaxies in three redshift bins of over ${0.2<z<0.8}$, 
and bins of spectroscopic galaxy stellar mass in the range ${9.5<\logM<12}$.
Unlike all previous conformity studies, Br16b did not utilize isolation criteria to select ``isolated primary'' or ``central'' galaxies, and therefore did not measure the same conformity statistic as in this work and other studies.
However, we can qualitatively compare our one-halo results to theirs.

For spectroscopic galaxies of stellar mass comparable our high-mass bin (${10.8 < \logM < 12}$) Br16b find a difference in the overdensity of quiescent galaxies only at ${0.4<z<0.6}$, while for spectroscopic galaxies of stellar mass comparable to our low-mass bin (${9.5 < \logM < 10.8}$) they find a difference in quiescent galaxy overdensity over the full redshift range they test at this mass:~${0.2<z<0.6}$.
This is qualitatively consistent with our one-halo results, although the \emph{magnitude} of the signal is much larger than our $\sim5$\% one-halo conformity signal.

\subsection{The Physical Driver of Two-Halo Conformity}

If large-scale tidal fields are the cause of two-halo conformity it should be possible to detect a correlation between the quenched fraction of central galaxies and large-scale environment.
In \S\ref{sec:environment} above we look for such a correlation in a sample of IP galaxies in the stellar mass range ${10.1 < \logM < 11.0}$ using the same methods at \citePB, and find the same trend that \citePB observed at low redshift in SDSS:~central galaxies are more likely to be quenched in overdense environments, \emph{independent of stellar mass}.
We also detect this correlation with greater significance than the typical measure of two-halo conformity described above in \S\ref{sec:signal}.

At a given stellar mass, large-scale environment evidently either \emph{does} impact central galaxy quenching or at least correlates with something that does.
While this correlation does not necessarily imply a causal connection, it is consistent with the He16 description of two-halo conformity being an indirect effect of large-scale tidal fields.
As such, it is likely observational evidence of assembly bias.

Interestingly, we also find that \emph{as long as a central galaxy is forming stars}, the efficiency of star formation does \emph{not} depend strongly (if at all) on large-scale environment.
\citet{Darvish16} find a similar result for a mass-complete sample of galaxies in the COSMOS field at $z\lesssim3$:~the median sSFR of \emph{star-forming} galaxies does \emph{not} vary significantly with environment, regardless of redshift and stellar mass.
However, because \cite{Darvish16} study all galaxies (not just central galaxies) their result is dominated by satellites in overdense environments.
We have shown that this result is true of just central galaxies as well.

Our result is \emph{not} solely due to the known relation between galaxy clusters and increased quenched fraction \citep[e.g.,][]{Cooper07}.
Roughly 8\% of the IPs in our sample reside in very overdense environments (i.e., they have $>30$ neighbors within $\sim4$ projected Mpc),
\emph{at all stellar masses} in the range we studied, and these IPs are not exclusively located in clusters;~they often lie along the large-scale filaments seen in Figure~\ref{fig:cone_diagrams}, as well as in more typical cluster-type environments.

Additionally, if our result that central galaxies are preferentially quenched in overdense environments were due to cluster-specific processes the magnitude of the effect should be greater for larger halo mass (and thus also for larger central galaxy stellar mass).
However, \citet{Weinmann06} found that the one-halo conformity signal in SDSS is independent of halo mass, and the two-halo SDSS signal found by K13 is \emph{stronger} at \emph{lower} stellar mass.

\subsection{The Importance of Large Survey Volume}\label{sec:large_volume}

As we have shown, cosmic variance dominates the uncertainty---and therefore the significance---of any conformity signal measured at intermediate to high redshift, due to the relatively small volume of sufficiently deep observational data currently available.
While a conformity signal in one or two small fields may be a robust measurement \emph{within that field}, we caution against drawing broad conclusions about any observed dependence of conformity on redshift or stellar mass from existing studies.
Simply put, more data are needed, and in particular, much larger volumes need to be surveyed with spectroscopic redshifts to faint depths in order to robustly test predictions of how conformity should evolve with cosmic time.

The Baryon Oscillation Spectroscopic Survey \citep[BOSS;][]{Dawson13} has obtained 1.5 million spectroscopic redshifts for luminous galaxies to $z\sim0.7$, but the mass distribution of this sample peaks at ${\mstar\sim10^{11.3}~\msun}$, and the sample contains almost no galaxies with stellar masses below ${10^{10.5}~\msun}$ \citep{Maraston13}.

The upcoming Dark Energy Spectroscopic Instrument \citep[DESI;][]{Flaugher14, Eisenstein15} survey is expected to obtain a 14,000~$\degsq$ nearly complete sample of $10^7$ bright (${r<19.5}$) galaxies, but only to $z\sim0.4$.  This will extend the current SDSS-type studies to $z=0.4$,
but deep, wide-area spectroscopic surveys are still needed at $z>0.4$ to test the theoretical predictions of Hearin et al.~and more accurately constrain galaxy evolution models across cosmic time.

The best current candidate for studying conformity at intermediate to high redshift is an upcoming survey with the Subaru Prime Focus Spectrograph \citep[PFS;][]{Takada14}.
This survey will observe 16~$\degsq$ of color-selected galaxies 
and AGN at $1<z<2$ to a depth of $J\simeq23.4$, obtaining a statistically complete sample 
of galaxies with stellar masses greater than $\sim10^{10}~\msun$ at $z\sim2$. 

%% file: conclusion.tex
\section{Conclusions}\label{sec:conclusion}

The existence of galactic conformity, or the observed correlation between 
the fraction of isolated, ``central'' galaxies that are quenched or have 
low sSFR and the fraction of neighboring ``satellite'' galaxies that are also
quenched, indicates that there is physics beyond the standard halo model 
of galaxy evolution.  In particular, whether a central galaxy ceases to 
form stars must depend on more than just the mass of the dark matter halo 
that that galaxy resides in.  

While the existence of galactic conformity was first measured ten years ago
in SDSS galaxies at $z<0.2$, it has only very recently been measured at 
intermediate and high redshift.  
Measurements of galactic conformity at higher redshifts is a very powerful 
tool for constraining how halo occupation models should move beyond the 
standard HOD model, and in particular for constraining what the quenching 
mechanism behind conformity must be.  

Previous measurements of conformity at $z>0.2$ relied on photometric 
redshifts; however, the large uncertainties of such measurements and the 
possibility of contamination of isolated galaxy samples using photometric
redshifts calls into question their usefulness.  These studies also only
probed so-called one-halo conformity, between central and satellite galaxies
within a given dark matter halo.  In SDSS there are clear indications that
conformity exists on larger scales, between halos (i.e., two-halo conformity).

Here we have tested for one- and two-halo galactic conformity at ${0.2<z<1}$ 
with a 5.5~$\degsq$ sample of $\sim60,000$ ${\mstar\gtrsim10^{9.3}~\msun}$ 
galaxies from PRIMUS, the largest existing spectroscopic redshift survey 
of faint galaxies to ${z\sim1}$.
Covering four spatially distinct fields, our sample allows us to probe a large 
cosmic volume and also account for the effect of cosmic variance on the conformity signal,
which we have shown can vary substantially between fields.

The primary conclusions of this work are:
\begin{enumerate}

\item
We detect a one-halo conformity signal at $3.6\sigma$, and a two-halo signal at $2.5\sigma$.  
The amplitude of the conformity signal is very small: only 5.3\% on one-halo 
scales and 1.1\% on two-halo scales.
Given the small size of the effect, it is critical to perform robust studies that take into account various possible 
systematic effects, including matching galaxy samples in both redshift and 
stellar mass, as well as using well-defined isolation criteria to identify central galaxies.

\item
That the conformity signal in PRIMUS is weaker than the signal observed in SDSS is consistent with the idea that galactic conformity is due to large-scale tidal fields, which predicts that the amplitude of the signal should decrease with increasing redshift.
This result is likely observational evidence for assembly bias.

\item
We observe a two-halo effect more robustly by measuring the star-forming fraction of central galaxies at fixed stellar mass as a function of large-scale environment, and find that central galaxies are more likely to be quenched in denser environments, independent of stellar mass.
Interestingly, the star formation efficiency of star-forming central galaxies does \emph{not} significantly decline in high-density environments.
However, environment does either help abruptly shut off star formation, or correlate with something that does.

\item
Ours is the largest area intermediate redshift conformity study to date, 
and the only measurement of conformity at $z>0.2$ performed with spectroscopic redshifts.
It is also the only detection of two-halo conformity at $z>0.2$.  
While our detections are robust, the fact that a survey the size of 
PRIMUS is only large enough to detect a $\sim$1\% two-halo conformity signal 
at the $2.5\sigma$ level illustrates the need for a next generation of 
deep, wide-field spectroscopic redshift surveys at $z>0.2$ to advance our understanding of galaxy and halo evolution.
Current predictions of the dependence of the strength of the conformity 
signal with mass and redshift to $z\sim1$--2 cannot be conclusively tested 
without spectroscopic data of comparable depth from additional fields.

\end{enumerate}